\documentclass[12pt,preprint]{aastex}

\usepackage{emulateapj5}

\newcommand{\et}{et al.}

\newcommand{\fv}{F_{var}}

\newcommand{\xte}{{\it RXTE}}

\newcommand{\ka}{K$_{\alpha}$}
\newcommand{\fka}{F$_{K\alpha}$}
\newcommand{\Msun}{\hbox{$\rm\thinspace M_{\odot}$}}

\slugcomment{}
\lefthead{Markowitz, Edelson \& Vaughan}
\righthead{Sy 1 X-ray Spectral Variability}

\received{2003 June 11}
\begin{document}
\title{Long-Term X-ray Spectral Variability in Seyfert~1 Galaxies}

\author{A.~Markowitz\altaffilmark{1}, R.~Edelson\altaffilmark{1},
S.~Vaughan\altaffilmark{2}}

\altaffiltext{1}{Dept.\ of Astronomy, Univ.\ of California, Los Angeles CA
90095-1562; agm@astro.ucla.edu, rae@astro.ucla.edu}
\altaffiltext{2}{Institute of Astronomy, Madingley Road, Cambridge, CB3~0HA, UK; sav@ast.cam.ac.uk}

\begin{abstract}

Direct time-resolved spectral fitting has been performed on
continuous {\it RXTE} monitoring of seven Seyfert~1 galaxies in order to
study their broadband spectral variability and Fe~\ka\ variability
characteristics on time scales of days to years.
Variability in the Fe~\ka\ line is 
not detected in some objects but is present in others,
e.g., in NGC~3516, NGC~4151 and NGC~5548 there are systematic 
decreases in line flux by factors of $\sim$2--5 over 3--4 years. The Fe~\ka\ line 
varies less strongly than the broadband
continuum, but, like the continuum, exhibits
stronger variability towards longer time scales.
Relatively less model-dependent broadband fractional variability amplitude ($\fv$) spectra
also show weaker line variability compared to the continuum variability.
Comparable systematic long-term decreases in the line and continuum
are present in NGC~5548.
Overall, however, there is no evidence for correlated variability
between the line and continuum,
severely challenging models in which the line tracks 
continuum variations modified only by a light-travel time
delay. Local effects such as the formation of an ionized skin
at the site of line emission may be relevant. 
The spectral fitting and $\fv$ spectra both 
support spectral softening as continuum flux increases.

\end{abstract}


\section{ Introduction }

There is substantial evidence to indicate that the central regions of Seyfert~1
Active Galactic Nuclei (AGN) consist of a geometrically-thin, 
optically-thick accretion disk
surrounding a supermassive (10$^{6-9}$$\Msun$) 
black hole. In the leading models, the inner disk is surrounded
by a hot, optically-thin corona that Compton-upscatters soft
photons thermally emitted from the disk to produce
an X-ray power-law continuum over the range 1--100~keV, though the 
geometry of the corona remains uncertain (e.g., 
Haardt, Maraschi \& Ghisellini 1994).
Furthermore, some optically thick material, such as the accretion disk,
intercepts and reprocesses some of the X-rays, producing the so-called 
Compton reflection hump above $\sim$10~keV and a strong Fe~\ka\ 
fluorescent emission line at 6.4~keV (Guilbert \& Rees 
1988, George \& Fabian 1991). Though the exact configuration
of accreting material throughout the central engine is not known,
the Fe~\ka\ line, as a primary signature of the
reprocessing, serves as a potential tool to trace  
the distribution of accreting gas.

Single-epoch spectral fitting has revealed a variety of profiles for
the Fe~\ka\ line. 
In some Seyfert~1 galaxies, the line is seen to be relativistically
broadened (e.g., FWZI up to 100,000~km~s$^{-1}$ in the well-studied 
broad line profile of MCG--6-30-15, Tanaka \et\ 1995)
and redshifted. This can be explained in terms of a
line originating in the innermost regions 
of the accretion disk, in close proximity to the black hole,
and affected by strong Doppler and gravitational redshifts (Fabian \et\ 1995).
However, the majority of Seyfert~1 galaxies observed with
{\it XMM-Newton} (Reeves 2002) and {\it Chandra} (Yaqoob, George \& Turner 2001)
feature a narrow Fe~\ka\ component, modelled to be 
only mildly broadened, with FWHMs of 
$<$10,000~km~s$^{-1}$ (e.g., Kaspi \et\ 2002,  
Nandra 2001, O'Brien \et\ 2001, Yaqoob \et\ 2001). For these objects,
a line origin in the outer accretion disk, putative molecular torus,
broad emission line region or narrow emission line region
may be more relevant. 

Studying the time-resolved variability of the line and continuum components
provides a complementary analysis to single-epoch spectral fitting.
For instance, Fe~\ka\ line emission originating 
in close proximity to the central black hole is expected to respond 
rapidly to changes in the illuminating continuum flux
(Stella 1990, Matt \& Perola 1992). The simplest models predict
that if the reprocessing time scale is negligible, then
variations in the Fe~\ka\ line should track continuum variations,
modified only by a time delay equal to the light-crossing time across
the inner disk. 
However, recent studies have not been able to support this picture.
Investigations into the behavior of the broad Fe~\ka\ line
in MCG--6-30-15 by
Iwasawa \et\ (1996, 1999), Lee \et\ (1999), Reynolds (2000),
Vaughan \& Edelson (2001, hereafter VE01), and Fabian \et\ (2002) 
have confirmed the presence of rapid Fe~\ka\ line variability 
in this object, but have found the line to vary considerably less
strongly than the continuum, and have found 
no correlation between continuum and line variations.
Yaqoob \et\ (1996) found the red wing of the Fe~\ka\ profile in NGC~7314 to
respond to rapid continuum variations, while the core did not respond as rapidly.
Conversely, Nandra \et\ (1999) found that the narrow core of the 
line in NGC~3516 tracked the continuum flux, while the broad 
wings displayed uncorrelated variability.
Petrucci \et\ (2002) used {\it XMM-Newton} to obtain two
observations of Mkn~841 separated by 15~hours;
while the continuum flux and shape
stayed nearly constant between the two observations, 
the narrow line virtually disappeared.

This paper aims to study the broadband spectral variability of
Seyfert~1 galaxies on time scales of days to years. This includes
quantifying the variability of the Fe~\ka\ line
and determining the relation between the line and the continuum flux.
To this end, samples have been assembled
that probe spectral variability on time scales of days to weeks
(hereafter referred to as ``short'' time scales) and time scales of 
months to years (hereafter referred to as ``long'' time scales), using
archival monitoring data from the {\it Rossi X-ray Timing Explorer (RXTE)}.
{\it RXTE} provides the largest effective area at 6~keV 
of any current X-ray instrument flown to date and thus offers 
the highest signal-to-noise of the Fe~K region of the X-ray spectrum.
The data reduction and sampling are described in $\S$2.
In order to track the line behavior,
direct time-resolved spectral fitting has been systematically performed
and fractional variability amplitude spectra have been constructed, presented in $\S$3.
The observational results are summarized in $\S$4.
The physical implications are discussed in $\S$5, followed by a brief
summary of the main results in $\S$6.

\section {Data collection and reduction}

The goal of this project was to obtain uniform samples 
to survey line and continuum variability behavior. 
Continuous {\it RXTE} archival 
monitoring data turning public before 2002 December,
when these analyses were performed, were used, as well as
the authors' proprietary observations of three Seyfert~1 
galaxies observed during Cycle~7.

All {\it RXTE} data considered in this paper were obtained with the
proportional counter array (PCA).
3--18~keV count rates were estimated with
the online {\sc W3PIMMS} tool, using mean
2--10~keV archival count rates and photon indices 
obtained either from previously published fits (e.g., Kaspi \et\ 2001)
or the Tartarus database.
The goal of time-resolved spectral fitting was to obtain 
at least ten high-quality spectra with at least 
$\sim$50,000 photons in each time bin;
sources that failed to meet this criterion were excluded.
This resulted in eight observations consisting of continuous monitoring
on short time scales
(e.g., sources observed once every 1--2 orbits for $\sim$1~ks
over a span of $\sim$30 days)
and an additional seven observations covering
long time scales (e.g., sources observed
once every 3 or 4 days for several years).
There were six objects common to both lists. This included
two MCG--6-30-15 observations on short time scales, 
an 8-day intensive scan henceforth denoted as short-term MCG--6-30-15~(I)
and a 53-day observation 
denoted as short-term MCG--6-30-15~(II).
This also included two MCG--6-30-15 observations on long time scales,
one spanning 326 days denoted as long-term MCG--6-30-15~(I)
and a 2093-day observation denoted as long-term 
MCG--6-30-15~(II). Additionally, 
NGC~7469 was monitored only on the short time scale.
Table~1 lists the date ranges and source parameter information.

The PCA consists of five identical
collimated proportional counter units (PCUs; Swank 1998).
For simplicity, data were collected only from those PCUs
that did not suffer from repeated breakdown during on-source time
(PCUs 0, 1, and 2 prior to 1998 December 23; PCUs 0 and 2 from 
1998~December~23 until~2000~May~12; PCU~2 only after 2000 May 12).
Count rates quoted in this paper are normalized to 1~PCU.
Only PCA STANDARD-2 data were considered.
The data were reduced using standard extraction methods and {\sc FTOOLS~v5.2}
software. Data were rejected if they were
gathered less than 10$\arcdeg$ from the Earth's limb, if
they were obtained within 30~min after the satellite's passage
through the SAA, if {\sc ELECTRON0}~$>$~0.1
({\sc ELECTRON2} after 2000~May~12),
or if the satellite's pointing offset was greater
than 0$\fdg$02.

As the PCA has no simultaneous background monitoring capability,
background data were estimated
by using {\sc pcabackest~v2.1e} to generate model files
based on the particle-induced
background, SAA activity, and the diffuse X-ray background.
This background subtraction is the
dominant source of systematic uncertainty (e.g., in total broadband count rate)
in \xte\ AGN monitoring data
(e.g., Edelson \& Nandra 1999).
Counts were extracted only from the topmost PCU layer to maximize
the signal-to-noise ratio.
All of the targets were faint ($<$~40~ct~s$^{-1}$~PCU$^{-1}$),
so the most recent versions of the 'L7-240' background models were used.
Source and background spectra were derived over the time intervals listed
in Table~1. These time bins were chosen to ensure adequate 
signal-to-noise in the Fe~\ka\ line
in the time-resolved fits, but also to
ensure adequate temporal resolution in determing the 
variability characteristics
of both the continuum and the line.
Response matrices were generated for each gain epoch
using {\sc pcarsp v.8.0}.

\section{Analysis}

The time-averaged fits, described in $\S$3.1, were performed with the goal of minimizing the
number of free parameters needed to produce an acceptable fit.
These model fits were then applied to the time-resolved data, as described in $\S$3.2. 
The errors on parameters derived from spectral fitting are discussed in $\S$3.3.
The time-resolved spectral fitting is discussed in $\S$3.4.
Continuum and line variability amplitudes are presented in $\S$3.5.
Temporal cross-correlation analysis is described in $\S$3.6.
Finally, the fractional variability amplitude ($\fv$) spectra are described in $\S$3.7. 
 
\subsection{Time-Averaged Spectral Fitting}

For all analyses, data below 3.5~keV were discarded in order to disregard
PCA calibration uncertainties below this energy and also to reduce the 
effects of warm absorbers in sources (e.g., MCG--6-30-15, Reynolds 1997; NGC~3783, Kaspi \et\ 2002;
NGC~5548, Nicastro \et\ 2000). Data above 18~keV were discarded as 
the source count rate became too low at higher energies
to obtain a good detection in each time bin.
All spectral fitting was done with XSPEC v.11.0.2 (Arnaud 1996).
Most long-term observations spanned multiple gain epochs,
requiring simultaneous fitting of data sets.

\setcounter{footnote}{0}

For all observations, the simple model of a power-law modified by 
cold absorption resulted in large residuals 
near 6~keV due to the Fe~\ka\ line; the 
addition of a emission component proved
significant at greater than 99.99$\%$ confidence in an F-test. 
Given the limit of PCA calibration 
(data/model residuals $\lesssim$2-3$\%$\footnote{See http://lheawww.gsfc.nasa.gov/users/keith/rossi2000/energy$\_$response.ps},
and the low energy resolution of the PCA ($\sim$1~keV at 6~keV), 
the PCA is largely insensitive to the detailed
shape of the Fe~\ka\ line. 
Both a single Gaussian and a Laor model component
(diskline model for a maximally rotating Kerr black hole; Laor 1991) 
were tested to parameterize the line emission. 
More complex fits such as double Gaussians or a Laor component plus a 
narrow Gaussian were either not required or did not yield sensible results;
the single-Gaussian and Laor fits alone gave sensible results.
For the Laor model component,
the inclination angle was kept fixed at 30$\arcdeg$, and the outer
radius kept fixed at 400 gravitational radii; in
no case did thawing either or both of
these parameters measurably improve the fits.
The emissivity index $\beta$ was constrained to fall in the range 0.0--9.9.
Laor models gave significantly better fits
(at $>$99$\%$ confidence in an F-test) than the single-Gaussian
parameterization in seven of the fifteen time-averaged spectra.
An implicit assumption in these fits is that the emission regions
in each source do not evolve substantially (i.e., major changes in
the geometry of the emitting regions) throughout the duration of the observation. 

For all sources except NGC~4151, neutral absorption was kept 
fixed at the Galactic value. For NGC~4151, 
modelling the column density with a value much higher than Galactic 
resulted in a significant ($>$99.99$\%$ confidence in an F-test) 
improvement in the model fit.
The column density values obtained, given the PCA calibration, 
were in reasonable agreement with each other and with the value of
4.6$\pm$0.2~$\times$~10$^{22}$cm$^{-1}$ in the spectral fit by
Schurch \& Warwick (2002).

For all model fits except the short time scale observation of 3C~390.3
and the Laor model fit to the short time scale observation of NGC~3516,
there were residuals above 10~keV that signalled the
need to incorporate a Compton reflection hump. 
Including a {\sc pexrav} component to model 
reflection of an underlying power law component
off optically thick and neutral
material (Magdziarz \& Zdziarski 1995) 
provided the necessary improvement to the fits, significant
at greater than 99$\%$ confidence compared to models with just a power-law
plus Fe~\ka\ line, modified by cold absorption. 
In the {\sc pexrav} component, the inclination angle of the reflector
was kept fixed at 30$\arcdeg$, 
and the power-law cutoff energy was kept fixed at 100~keV. 
All element abundances were kept fixed at solar; as noted by
VE01 a change in Fe abundance caused the absolute values of 
spectral fit parameters to change slightly,
but the variability pattern, the focus of this paper, remained unchanged.

The basic model thus used, a power-law component plus
{\sc pexrav} (where applicable) plus a parameterization
for the Fe~\ka\ line as described above, all modified by neutral 
absorption, gave
reasonable fits to the data. The exceptional signal-to-noise in these
time-averaged spectra invariably meant that systematic errors
associated with the background modelling and spectral 
response of the PCA dominated the uncertainties.
Consequently, finding a statistically acceptable time-averaged fit
(with $\chi_{\nu}^2$~$\sim$~1) was unlikely even if the model
was representative of the instrinsic spectrum of the source.
The data/model residuals seen were consistent with the
limit of the PCA calibration.
Tables~2--3 summarize the 
time-averaged fitting results for both the single-Gaussian and 
Laor models, respectively, on both time scales.

While higher-resolution X-ray instruments can detect one or both of
the Fe~\ka\ broad and narrow components, the PCA is sensitive only
to the combined flux of the two components. Recent 
{\it XMM-Newton} model fits can provide insight 
as to which component, the broad or narrow, dominates the \xte\
view in each object. ({\it Chandra-HETG} can additionally constrain the narrow component but
is mostly insensitive to the broad component.)
For instance, the equivalent width of the broad component has been measured
to be much greater
than that of the narrow component in both MCG--6-30-15 (Wilms \et\ 2001) 
and NGC~3516 (Turner \et\ 2002). The present \xte\ fits also indicated 
that the Fe~\ka\ lines in these two objects are among the broadest of the sample,
consistent with the bulk of the line emission originating in the inner 
accretion disk. In constrast, the narrow component seems to dominate in
NGC~4151 (Schurch \et\ 2002), NGC~5548 (Pounds \et\ 2003, Yaqoob
\et\ 2001) and NGC~7469 (Blustin \et\ 2003), with a small upper limit
or no detection of the broad component. The present \xte\ fits
for NGC~4151, for instance, indicated the narrowest line
of the sample, consistent with emission far from the central black hole.
The equivalent widths of the broad and narrow components
in NGC~3783 are approximately equal (Blustin \et\ 2002), indicating
that emission from both the inner accretion disk and a more distant location
is relevant in this source.

\subsection{Time-Resolved Spectral Fit Components}

The above models were then applied to the time-resolved data.
Spectral fits to time bins spanning gain epoch changes 
required simultaneous fitting of the two data sets.
The Fe~\ka\ line was always detected in each time bin, as the addition of 
a line emission component was significant at $>$99$\%$ confidence in an F-test
for nearly every time bin (in the single exception, the last 
time bin of the long time scale
observation of NGC~5548, it was significant only at 98.1$\%$ confidence).
While fitting the time-averaged models to their
respective time-resolved spectra, 
it was not practical to thaw parameters such as the Fe~\ka\ 
line profile shape (line width in the single-Gaussian
models; inner radius and emissivity index $\beta$ 
in the Laor models) or line peak energy, given the PCA resolution and sensitivity.
The F-test was used to determine which free parameters to include in the fit.
The time-resolved spectral fits were repeated with each of photon index
$\Gamma$, reflection fraction
$R$ (ratio of the reflection normalization at 1~keV to the normalization
of the power law) and line flux normalization \fka\
frozen at their time-averaged values, and the values of
total $\chi^2$ were compared via an F-test to determine which parameters
could justifiably be allowed to be free and which
should be kept fixed in the fitting, as listed in Table~4.
There was no formal significant evidence
($>$99.7$\%$ confidence) of the need to thaw $R$ in any object.
For most sources (6/8 short time scale and 6/7 long time scale
observations), there was significant ($>$99.7$\%$ confidence) 
evidence of the need to thaw $\Gamma$.
There was also significant ($>$99.7$\%$ confidence) 
evidence of the need to thaw \fka\ in the
short time scale observations of MCG--6-30-15~(I), NGC~3516 and NGC~7469
as well as in the long time scale observations of
MCG--6-30-15~(II), NGC~3516 and NGC~4151.
These results initially suggest statistically significant variations 
in the line in at least some of these objects.
Table~5 lists the mean values of the 2--10~keV flux
$F_{2-10}$, $\Gamma$ and \fka\ derived
from the time-resolved spectral fitting. 
$F_{2-10}$ was measured from the model fits.

\subsection{Error Estimation}

For the majority of the individual time-resolved fits, the
resulting values of the fit statistic $\chi^2_\nu$ were rather low
($\chi^2_\nu$ $<$ 1.0 for 243 of the 316 total time bins). 
As discussed by Nandra \et\ (2000) and VE01, this was 
most likely due to the standard software's 
overestimation of the errors in any given energy bin in the
net spectrum. These errors are produced by 
combining the errors in the source with those of the modelled 
background. The software uses Poisson statistics, estimated from
a large quantity of background data, to estimate the background 
error. For short exposures, this results in an overestimate
of background errors, which should typically be negligible
compared to the error on the total (source plus background)
rate. To better ascertain the variability of the measured
spectral parameters, the standard
errors produced by the software were discarded and new 1$\sigma$ errors were estimated 
based on the variability properties of the derived light curves.
The point-to-point variance was used; this is
defined as  \[ \sigma^2_{PP} = \frac{1}{2} \frac{1}{N - 1} \sum_{i=1}^{N-1}(x_i - x_{i+1})^2 \] where $x_i$ are each of the
N total points in the light curve. VE01 used simulations to validate the
accuracy of these errors.
These are conservative error estimates;
only trends in the light curves sustained over several time bins
will exceed the expected error using this formulation.
These errors are listed in Table~5 for $\Gamma$
and $F_{K\alpha}$ and used hereafter.
Errors on values of $F_{2-10}$ within a time bin 
were determined from the 2--10~keV continuum light curve,
using the mean flux error 
on the $\sim$1~ks exposures in that time bin.

\subsection{Time-Resolved Broadband Spectral Fits}

Figures 1 and 2 show light curves of $F_{2-10}$, $\Gamma$, and \fka\
for the single-Gaussian model on short and long time scales,
respectively. The results of the Laor model time-resolved 
fits showed identical variability 
patterns and nearly identical time-averaged values to
the single-Gaussian fits, and they are not plotted.

This spectral fitting was performed keeping the value of $R$ 
fixed, i.e. with the normalization of the {\sc pexrav} reflection component
varying in sync with the normalization of the power-law.
This was done as there was no formal statistical requirement to allow
the value of $R$ to vary in the fits; freezing $R$
also minimized the influence
of this free parameter on measurement of $\Gamma$ and F$_{K\alpha}$.
However, as the reflection
continuum and Fe~\ka\ line are likely to be produced in the same location
(e.g., the accretion disk) it would perhaps be expected that the
reflection component tracks variations in the line rather than the
power-law continuum. Time-resolved spectral
fitting was repeated with the normalization of the reflector tied to the \fka\
instead. The resulting light curves of \fka\ 
were essentially the same as those with a fixed value of $R$.

\subsection{Fractional Variability Amplitudes}

Fractional variability amplitudes 
($\fv$; e.g., Vaughan \et\ 2003, Edelson \et\ 2002)
were measured to quantify the instrinsic variability amplitude
relative to the mean count rate and 
in excess of the measurement noise in the continuum and line light curves;
\begin{equation}
\fv = \sqrt{S^2 - \langle \sigma_{err}^2 \rangle \over \langle X
\rangle^2},
\end{equation}
where $S^2$ is the total variance of the light curve, $\langle
\sigma_{err}^2 \rangle $ is the mean square error and $ \langle X \rangle
$ is the mean count rate of $N$ total points.
The error on $\fv$ is
\begin{equation}
\sigma_{\fv} = \sqrt{  \left\{ \sqrt \frac{ \langle\sigma_{err}^2\rangle}{N} \cdot \frac{1}{\langle X \rangle} \right\} ^2         +       \left\{ \sqrt \frac{1}{2N} \cdot \frac {  \langle\sigma_{err}^2\rangle }{\langle X \rangle^2 \fv} \right\} ^2             }
\end{equation}
as discussed in Vaughan \et\ (2003). 
This error formulation estimates
$\sigma_{\fv}$ based on 
random errors in the data itself, and not
due to random variations associated with red-noise processes.
Furthermore, these estimates are only an approximation to the true
values of $\sigma_{\fv}$
since the data errors used here 
instead have been estimated from the point-to-point variance.
Table~6 lists the continuum and line variability amplitudes.

\subsection{Cross Correlation Analysis}

Temporal cross-correlation functions were calculated for
F$_{2-10}$--F$_{K\alpha}$, F$_{2-10}$--$\Gamma$, and  $\Gamma$--F$_{K\alpha}$
using both the discrete correlation Function
DCF; Edelson \& Krolik 1988) and the interpolated correlation
function (ICF, White \& Peterson 1994). ICF lag uncertainties
were calculated using the bootstrap method of Peterson \et\ (1998); however,
this method often underestimates the true errors 
introduced by uneven sampling, e.g., Edelson \et\ (2001).  
Figures 3 and 4 and Tables 7--9 show
the cross-correlation results, where a positive shift denotes the first 
quantity leading the second. The DCF and ICF time bins used were
1.0 and 0.5 times the time bin size in the time-resolved 
fits, respectively.

One must be skeptical of cross-correlation analysis in these circumstances,
however, due to the red-noise nature of the continuum
and line fluxes (as speculated below).
Cross-correlation between two unrelated red-noise
noise light curves can randomly yield spurious correlations
with higher than expected values of the
correlation strength $r_{max}$ (e.g., Welsh 1999). For instance,
the $r_{max}$~=~0.7 optical-to-X-ray
lag in NGC~3516 claimed by Maoz \et\ (2000) was not 
confirmed with additional data by Maoz \et\ (2002).
Combined with the fact that most of these light curves only have 10--20 points,
spurious correlations and anticorrelations, including those 
based on very few points (i.e., trends much shorter than the
total duration), can arise. 
Discarding peak lags greater than $\sim$1/3 of the total duration
can eliminate some of these spurious lags; 
this has been done for the values reported in Tables~7--9
and the functions plotted in Figures~3 and 4.
However, as has been done in the present work, it is more practical 
to not trust any lag where $r_{max}$ is not very close to 1.0,
applying relatively more skepticism for lower values of $r_{max}$.

Cross-correlation analysis of F$_{2-10}$ and F$_{K\alpha}$
in the short time scale observations yields
peak correlation coefficients in the range 0.17--0.72.
These values do not readily convince us that the analysis has yielded
'believable' correlations. For instance,
the highest correlation coefficient, $r_{max}$~=~0.72,
arises in NGC~3783, which
only has 12 points and for which substantial
variability in the line is 
not detected (see Table~4). 
NGC~5548, for which cross-correlation analysis yields the second-highest
value of $r_{max}$ in the short time scale sample, is a similar situation:
the reported lag is far from zero, based on only a handful of points, 
and one cannot conclude that this is a reliable correlation robustly detected
in excess of the long time scale red-noise variations present.
It follows that the lag delays and lag errors listed in Table~7
must also be regarded with a large dose of skepticism.

For long time scale F$_{2-10}$--F$_{K\alpha}$ correlations,
the values of $r_{max}$ lie in the range 0.45--0.87.
Despite the presence of higher values of   
$r_{max}$, however, caution is again warranted.
For instance, as the continuum and line light curves are both
dominated by large decreases in NGC~3516 and NGC~5548,
the resulting cross-correlation functions have moderately-high values of
the correlation
coefficient at nearly all lags tested, and the peak correlation coefficients
do not provide evidence of a strong correlation ``in excess'' of this 
artefact.
If there were an increase in the values of $r_{max}$ towards
a particular time scale, this would indicate
more of a physical relevance
associated with that time scale.
However, only three objects show an increase in 
their respective values of average DCF and ICF $r_{max}$
towards longer time scales. 

$\Gamma$--F$_{K\alpha}$ cross-correlation functions show similar correlation
coefficient values as the F$_{2-10}$--F$_{K\alpha}$ cross-correlation functions.
However, many cross-correlation functions between F$_{2-10}$ and
$\Gamma$ on both time scales
yield higher values of $r_{max}$. 
The values of $r_{max}$ on short time scales
span 0.54--0.92; those on long time scales
span 0.31--0.94 (0.62--0.94 if NGC~4151 is excluded).
All but one of the lag delays are consistent with zero.

\subsection{$\fv$ Spectra}

$\fv$ spectra, measuring $\fv$ of the 
net (background-subtracted) count rate in each channel, were
calculated as a relatively less model-dependent assessment of the
broadband spectral variability. Such
analysis allows one to study the behavior of the varying components.
Source and background spectra were derived over the time intervals listed
in Table~10 with the goal of constructing light curves of the flux 
in each channel for at least $\sim$20 points. 
In addition to the observations used in the above analysis,
$\fv$ spectra were measured for an additional four objects on short time scales
and five objects on long time scales; the total of twelve
observations on each time scale included nine objects common to both
lists.
Because the PCU gain settings changed three times since launch and
the channel energy boundaries differ between gain epochs,
the longest duration that did not cross a gain epoch boundary was used.
The error on each light curve point was assigned using the point-to-point
variance (see $\S$3.3).
The highest-energy bins were re-grouped as necessary to insure
adequate signal-to-noise
in all energy bins.
For the soft-spectrum ($\Gamma$~$\sim$2.7; Vaughan \et\ 1999) source 
Akn~564, data above 9~keV were discarded
to minimize the influence of systematic variability in the {\it RXTE}
background models.
Similarly, in Fairall~9
($\Gamma$~$\sim$~2.2 over 2--10~keV, TARTARUS database), data above 
14.3~keV were discarded for the short time scale data.
$\fv$ and $\sigma_{\fv}$ were calculated in each channel using 
the equations in $\S$3.5.
The final $\fv$ spectra
are presented in Figures 5 and 6 for short and long time scales respectively.
For the seven sources with published power spectral densities 
(Markowitz \et\ 2003, Uttley, McHardy \& Papadakis 2002),
the relative continuum variability amplitudes obtained on short and long time scales are in
reasonable agreement 
with values predicted from their published power spectral density shapes. 


\section{Results}

A summary of the observed continuum broadband flux and spectral variability properties of the sample is given in $\S$4.1.
The observed line variability properties are summarized in $\S$4.2. Finally, a comparison between 
the continuum and line variability properties is given in $\S$4.3. Notes on individual sources are deferred to Appendix~A.

\subsection{Continuum variability}

All objects display significant variability in continuum flux:
in $\chi^2$-tests against the constant hypothesis, the constant hypothesis is rejected at $>$99.99$\%$ confidence
for all continuum light curves. Visual inspection of Figures~1 and 2 confirms this.
There is evidence for the continuum to vary more strongly towards longer time scales: of the six objects with both 
short- and long time scale monitoring, the value of continuum $\fv$ increases towards longer time 
scales for five
objects. The $\fv$ spectra also exhibit stronger continuum
variability on longer time scales in most sources. Both these analyses support an increase in 
variability amplitude towards longer time scales 
that is consistent with the long-established red-noise nature of
Seyfert~1 continuum X-ray emission. 

The majority of $\fv$ spectra show the softer energies to be the most variable, consistent with softening of 
the dominating broadband X-ray continuum as sources brighten (e.g., Markowitz \& Edelson 2001, Nandra \et\ 1997).
The only exception is Akn~564, which exhibits a much flatter dependence on $\fv$ with energy.
The presence of F$_{2-10}$--$\Gamma$ correlations with no measureable lags  
in many objects on both time scales supports this idea. Such behavior has been 
observed numerous times before (e.g., VE01, Done \et\ 2000, Lee \et\ 2000, Chiang \et\ 1999, Nandra, Pounds \& Stewart 1990). 

Above $\sim$10~keV, unmodelled systematic variations in the background can 
contribute greatly to the total observed variability.
In most sources, therefore, the influence of the variability of the 
Compton reflection hump on the broadband continuum above $\sim$10~keV cannot be reliably constrained from 
the $\fv$ spectra. 

\subsection{Fe \ka\ line variability}

Given that the 5--7~keV continuum flux greatly outshines the line flux
and given the nature of spectral fitting, it is possible that there could exist some
systematic effect that corrupts the determination of the continuum level at the 
$\sim$few per cent level, in turn corrupting determination of the line flux by
$\sim$10 percent or higher. However, such an effect is not readily identifiable,
and the line variations measured are hence assumed
intrinsic to the sources.

Time-resolved spectral fitting has 
revealed a wide range of variability in the line 
on both time scales. 
On short time scales, MCG--6-30-15~(I) shows evidence for line variability: a $\chi^2$-test 
against a constant hypothesis results in rejection of that hypothesis at $>$99$\%$ confidence.
Similarly, the constant hypothesis is rejected in MCG--6-30-15~(II) and NGC~3516, 
but only at 98$\%$ and 95$\%$ confidence, respectively. 

On long time scales, the number of objects showing line variability increases:
in three of the seven observations (NGC~3516, NGC~4151 and NGC~5548), the constant line flux hypothesis 
is rejected at $>$99.99$\%$ confidence.
In two additional observations, MCG--6-30-15~(I) and 3C~390.3, the constant line flux
hypothesis is rejected, but only at 97.9$\%$ and 97.7$\%$ confidence, respectively. The most noteable trends in line flux
are the gradual decreasing trends throughout the long time scale observations of NGC~3516, NGC~4151 and NGC~5548. 
Lack of line variability cannot be ruled out in MCG--6-30-15~(II) and NGC~3783.
Overall, one can conclude that Fe~\ka\ line variability is more commonly detected towards longer time scales.

This tendency is confirmed using measured values of $\fv$: 
the value of line $\fv$ increases towards longer time scales for four objects, the values of 
line $\fv$ on both time scales are consistent for the remaining two objects, and in no case does line 
$\fv$ decrease towards longer time scales. It is therefore
plausible to conclude that, at least on the time scales probed herein, the iron line variability is,
like the continuum variability, a red-noise process.

The average maximum peak-to-trough line flux variation 
        detected is a factor of $\sim$2 for all observations; typically, 
        this is a factor of $\sim$4.5 greater than the typical error bar.
        The largest peak-to-trough variations are $\sim$3, as in the short-term
        observation MCG--6-30-15 (II) and the long-term observations of
        NGC~4151 and NGC~5548.
        The smallest peak-to-trough variations are $\sim$40--50\%, as with the
        short-term observations of NGC~3516 and NGC~3783 and the
        long-term obsrvations MCG--6-30-15 (II) and NGC~3783; typically this is
        a factor of $\sim$2 greater than the typical error bar. With the 
        time bins chosen in this experiment to maximize the 
        variability-to-noise in the iron
	line light curves, this experiment may not be sensitive to
	smaller amplitudes of variability (peak-to-trough variations of $\sim$30$\%$ or less).

\subsection{Relating continuum and line variability}

Visual examination of the continuum and line light curves does not reveal any obvious relation
between them in most objects. Large-amplitude trends in continuum often go unmatched by simultaneous
changes in the line; prominent examples of this are listed for individual sources below.
In only a few observations do the continuum and line appear to track each other: in the long time scale
observation of NGC~5548, both the continuum and line decrease by a factor of $\sim$3 over the 4-year duration.
There are also somewhat similar decreasing trends dominating both the continuum and line
in the long-time scale observations of NGC~3516 and NGC~4151.

Measured variability amplitudes, $\fv$ spectra, and the increased detection of line variability in the light curves towards 
longer time scales all indicate that both the continuum and line appear to be more strongly variable towards 
longer time scales in most objects.
However, there is evidence that the line does not
        vary as strongly as the continuum. In 11/15
        observations, the value of continuum $\fv$ is greater than the
        line $\fv$.  For the eight of these eleven observations
        with a defined value of line $\fv$ (positive excess variance),
        the continuum and line excess variances are compared via an F-test.
        In six of these eight observations, the null hypothesis of the 
        continuum and line excess variances being
	consistent is rejected at $>$98.2$\%$ confidence.
        In four of the fifteen observations, the values of 
        continuum and line $\fv$
        are consistent with each other. For three of these
        observations, an F-test cannot reject the null hypothesis of the  
        continuum and line excess variances being consistent at greater 
        than 79$\%$ confidence; it is rejected at $>$99.99$\%$ confidence
        in the fourth observation (the long-term NGC~3783 observation).
        There are no observations with a larger value of $\fv$ for the
        line compared to the continuum.

        As seen in Figures 5 \& 6, in many of the $\fv$ spectra,
        there is a reduction in variability at $\sim$6 keV.
        The
	average drop in $\fv$ at $\sim$6~keV relative to the surrounding
	continuum is $\sim$10$\%$, 
	or $\sim$5 times larger than the typical errors.
	Best fit lines to the $\fv$ spectra 
	including and excluding the $\sim$5--7 keV region
	indicate that the dip is significant 
	at greater than 90$\%$ confidence in an F-test for 11/24
	observations. These dips are consistent with dilution of the 
        $\sim$5--7~keV 
	continuum variability by relatively reduced line variability. 
	Such a feature has been noted previously in the $\fv$ spectra of 
        NGC~4151 
	(Schurch \& Warwick 2002) and MCG--6-30-15 (Fabian \et\ 2002).
	The presence of this feature supports the argument for decoupling of 
	the line and continuum variability properties. Additionally, 
        cross-correlation analysis does not reveal any significant 
        temporal correlations between continuum and line variations.

Overall, there hence is little support for a simple or direct temporal relation between the 
observed continuum and Fe \ka\ line variability properties. 
This result broadly agrees with the study of Weaver, Gelbord \& Yaqoob (2001), who did not find any 
evidence that Fe~\ka\ line variability on time scales of days to years was 
correlated with continuum variations; their {\it ASCA} 
sample included all of the objects in this paper's time-resolved 
spectroscopy samples except 3C~390.3, though with much less continuous sampling.

\section{Discussion}

This paper examines the variability properties of both the continuum
and Fe \ka\ line on time scales from days to years.
The continuum variability properties observed in the current sample
have been observed before in many Seyfert~1 galaxies.
The data show an increase in broadband variability amplitude towards 
longer time scales, demonstrating the red-noise nature of the continuum variability.
Additionally, there is an increase in variability amplitude 
towards the softer X-rays, consistent with sources' softening as they brighten.
In contrast, the Fe \ka\ line shows a divergence of behaviors.
Most observations show no clear trend whatsoever.
Only a few observations show temporal trends in the line flux that mimic the continuum to at 
least some degree, and only on time scales of years. 
As discussed in $\S$4.3, however,
        the line varies less strongly than the continuum in many 
        sources, as seen from a comparison of the spectral fit line and continuum
        flux light curve variability amplitudes 
        as well as from a reduction in broadband $\fv$ near 6 keV seen in many 
        sources' $\fv$ spectra. The relationship between continuum and line 
        variability is examined further in $\S$5.1. The broadband spectral 
        variability results are discussed in $\S$5.2.

\subsection{Fe~\ka\ line variability}

This paper has demonstrated the presence of a range of variability behavior in the Fe \ka\ line. 
The results permit tests of simple models in which the bulk of the continuum originates in a central 
corona and the line variations are driven solely by continuum flux fluctuations modified by 
light-travel time effects. For instance, such models imply that there should exist a connection between 
the origin of the line emission (which can be inferred from the resolved profile shape) and the presence 
of rapid line variability. If the bulk of the line is emitted in the inner accretion disk
(as implied by the broad line profiles in MCG--6-30-15 and NGC~3516), then the continuum and line  
variations should be correlated on time scales comparable to the light crossing time of the inner accretion disk. 
This is because the reprocessing times are likely very short compared to the light travel times, which dominate
the correlation. 
However, the time bins used in this paper are typically at least
      two orders of magnitude longer than the expected light crossing
      time scale of the putative inner accretion disk.  This means that any
      variations on this time scale would be grossly undersampled.

If the bulk of the line is emitted far from the central corona (e.g., as implied by the narrow line profiles
for NGC~4151 and NGC~5548), then one would expect the line variations to track the continuum variations
only on sufficiently long time scales (e.g., longer than days--weeks if the line is emitted in a region coincident 
with the broad line region; Kaspi \et\ 2000). For instance, the presence of similar long-term trends
in the continuum and line in NGC~5548 supports a model in which the bulk of the line 
emission originates far from the central corona in this object. More specifically,
it is consistent with the prediction of Yaqoob \et\ (2001) that the flux of the narrow core in NGC~5548, estimated to 
originate in a region consistent with the broad line region in this object, 
should track the continuum flux on time scales of $\sim$5~--~$\sim$100 days.
However, such trends are not ubiquitous.
For Fe~\ka\ lines originating far from the central corona, one would expect to see an increase
in number or strength of continuum--line correlations as one goes
towards towards long time scales. However, there is no evidence from the current sample to support this.

The lack of correlation between the continuum and the line suggests that modification to this 
straightforward picture is required. 
One possible explanation for the reduced variability in the line relative to the continuum
is that a portion of line emission originating in the inner disk is absent or suppressed.
The inner portions of the accretion disk may be 
truncated, highly ionized, or radiatively inefficient (e.g., as in an advection-dominated accretion flow; Narayan \& Yi 1994).
Geometry may play a role: a relatively more 
concave disk, for instance, can reprocess a larger fraction of 
continuum photons in its outer regions (Blackman 1999).
For cases in which the bulk of the observed line flux originates in the outer disk,
the direct response of the line peak to rapid continuum variations is reduced.
Another possibility is that if the X-ray continuum emission is produced in a 
corotating flare some height above the reflecting disk, then the orbital motions of 
the flare and disk can also lead to complex temporal behavior in the observed line flux
(Ruszkowski 2000).
Fabian \& Vaughan (2003) suggested that light-bending near the
black hole can lead to substantial de-coupling of the observed
continuum and line fluxes. This happens
because much of the emission from the X-ray source is
focussed on to the disc whereas only a small fraction is able to
reach the observer. In such a scenario small changes in the
geometry of the X-ray source can lead to relatively large
changes in the amount of continuum radiation reaching the
observer (hence the strong continuum variations) but relatively
little change in the total amount of radiation illuminating the
disk.


Another intriguing possibility relies on the presence of an ionized skin on the disk surface. Thermal instabilities  
can lead to formation of such an ionized skin (Ross, Fabian \& Young 1999; Nayakshin, Kazanas \& Kallman 2000) with most 
of the observed neutral line emission originating in the colder layers beneath. The ionized skin can scatter incoming 
continuum photons and outgoing line photons, affecting the observed  
neutral line flux, as discussed by Nayakshin (2000) and Nayakshin \& Kallman (2001). 
However, changes in the ionization state of the skin are linked to the dynamic time scale for
readjustment of hydrostatic equilibrium (Nayakshin \& Kazanas 2002 for a lamppost illuminating model; see also Collin \et\ 2003 
for a multiple-flare illuminating model). This time scale is usually an order of magnitude or more
larger than the light-crossing time; this can be $\gtrsim$minutes--hours if most of the line emission
originates in the inner regions of a typical thin accretion disk (e.g., around a 10$^{6-8}$~$\Msun$ black hole).
The observed line flux hence may not respond to continuum variations shorter than this time scale,
resulting in decorrelation of the observed continuum and line flux variations.
Furthermore, the ionization state of the skin depends on the high-energy cutoff of the power-law continuum
(Nayakshin \& Kallman 2001); this could be variable, but it is difficult to constrain with {\it RXTE}. 
Interestingly enough, such a variation in this parameter could lead to a change in Fe~\ka\ line flux even without
a change in 2--10~keV continuum flux.
However, recent studies have not been able to directly link the line flux behavior
with the ionization state of the disk, e.g., Ballantyne, Vaughan \& Fabian (2003).



All of the above scenarios are still consistent with the prediction that both broad and narrow 
Fe~\ka\ lines should eventually 
track continuum variations on time scales of months to years. The finding that the line 
decreased along with the continuum in the long time scale observation of NGC~5548 is 
consistent with this idea; however, such correlated behavior is not seen throughout the entire sample.
For line emission originating in the outer disk or molecular torus, 
ionization in the reflecting material is negligible due to the low X-ray continuum flux.   
However, other processes may be relevant for decorrelating long-term continuum and line variations.
For instance, it may be possible for an ionized wind (e.g., George \et\ 1998)  
near the site of reflection to mimic an ionized 'skin' in these regions.

\subsection{Broadband Spectral Variability}

Most of these objects show a strong correlation with no measureable lag between 2--10~keV continuum 
flux and photon index. Additionally, $\fv$ spectra show that in nearly all the objects, the soft-spectrum 
(narrow line) source Akn 564 being an exception, variability amplitudes tend to increase towards relatively 
softer energies. These observations show that the broadband emission steepens as it brightens, in line 
with numerous similar observations (e.g., Magdziarz \et\ 1998, Done \et\ 2000).

The time-resolved spectral fitting of MCG--6-30-15 performed from {\it ASCA} data by 
Shih, Iwasawa \& Fabian (2002) and from {\it XMM-Newton} data by Fabian \& Vaughan (2003) 
point towards a scenario in which flux-correlated changes in the overall 2--10~keV continuum slope are explained by the 
superposition of a soft, variable power-law likely associated with coronal emission
and a harder, much less variable spectral component likely associated with
the Compton reflection hump.  Taylor, Uttley \& McHardy (2003) use flux--flux plots to show that such a
description of spectral variability works in MCG--6-30-15 and NGC~3516.
Specifically, in this model, the intrinsic slope of the underlying power-law does not vary
significantly; rather, it is the {\it relative contributions} of the two spectral
components that change the overall spectral shape as observed.
This model does not require physical properties of the power-law emitting corona 
such as electron temperature, optical depth, or flare size to vary significantly.
Furthermore, this two-component model may have relevance to explaining the observed Fe~\ka\ variability.
As the Fe~\ka\ line and Compton hump are expected to
originate together, a relatively constant Compton hump
could be associated with variability in the Fe~\ka\ line
that is reduced relative to the continuum variability, as observed
(e.g., Shih \et\ 2002). 


Alternate models to explain the observed spectral variability rely on flux-correlated changes in 
the power-law slope of the coronal component.
In the context of popular thermal Comptonization corona 
models (Haardt, Maraschi \& Ghisellini 1997, Poutanen \& Fabian 1999), a flux--photon index
correlation can arise if the X-ray variability is due to changes
in the input seed photon population. In contrast to the two-component model above, 
some physical property of the corona may vary with
the 2--10~keV X-ray flux in these types of models.

The $\fv$ spectra of the soft-spectrum (narrow-line) Seyfert~1 galaxy Akn~564 display a much flatter dependence 
on energy in the 3.5--9~keV bandpass (and may even be slightly
increasing with energy on the long time scales) relative to
the other objects, most of which are hard-spectrum (broad-line) Seyfert~1 galaxies.
This has been seen previously in Akn~564 by Edelson \et\ (2002) as well as in
other soft-spectrum sources (Vaughan \et\ 2002).
This could imply that, in the context of the spectral-pivoting model above, 
the broadband spectral shape changes very little
in soft-spectrum sources, perhaps due to the lack of variation of some physical parameter
that does vary in broad-line Seyfert~1 galaxies.
Alternatively, if the two-component model above is relevant, it is possible that 
soft-spectrum sources have much weaker hard/constant components, leading to relatively little change
in the overall spectral shape with flux.
Overall, however, it is difficult to present a simple, unified phemonological model
to simultaneously explain the dependence on energy of the
variability in broad-line Seyfert~1 galaxies and lack thereof in narrow-line Seyfert~1 galaxies.




\section{Conclusions}

Time-resolved broadband spectral fitting has been systematically performed
on near-continuous \xte\ monitoring data of
seven Seyfert~1 galaxies in order to probe the
behavior of emission components on time scales of
both $\sim$days to $\sim$weeks and $\sim$months to $\sim$years. 
Fe \ka\ line variability is present at a variety of levels. It is readily 
        detectable in some observations; for instance, the average peak-to-trough
        variations are a factor of $\sim$2 and as high as $\sim$3 in three cases
        (the short-term observation MCG--6-30-15 (II) and the long-term observations of
        NGC~4151 and NGC~5548). In other observations, the line does not display appreciably
        variability; peak-to-trough variations are only $\sim$40--50$\%$. 
        The measured values of $\fv$ indicate that a higher fraction of the objects sampled
        show detectable line variability towards longer time scales.  
Of particular interest are systematic
long-term decreases in line flux over durations of 3--4 years in NGC~3516,
NGC~4151, and most noteably NGC~5548.
Furthermore, the fractional variability amplitude of the line 
increases towards longer time scales in many objects, ``tracking'' the
red-noise continuum.
However, fractional variability amplitudes show that the line does not 
vary as strongly as the continuum. $\fv$ spectra also generally show a dilution in $\sim$5--7~keV
continuum variability due to the relatively less variable Fe~\ka\ line.
In NGC~5548, the continuum and line simultaneously
decrease by similar factors over a 4-year period; the long-term observations
of NGC~3516 and NGC~4151 also show somewhat similar trends in continuum and line flux.
For the sample as a whole, however, there appears to be no simple temporal
relation between the continuum and the line, contradicting simple models
in which the line flux is driven solely by continuum variations.
Local effects such as an ionized skin
may be relevant in decorrelating the continuum and line variations,
particularly for lines originating in the inner accretion disk.
The strongest statement that can be made on the basis of these
results is that for most of the Seyfert~1 galaxies studied, the Fe~\ka\ emission
is less variable than the continuum, and in those cases where
there does appear to be evidence for changes in the line flux,
these changes are not well correlated with the continuum.

There is evidence for a correlation with no measureable lag
between continuum flux and measured photon index on both time scales in many sources.
Relatively less model-dependent
$\fv$ spectra generally show the broadband continuua 
to be more variable at relatively softer energies, supporting 
Seyfert~1 galaxies' softening as they brighten, though the soft-spectrum source Akn~564,
with its flat $\fv$ spectra, is an exception.

\acknowledgments
The authors acknowledge the dedication of the entire \xte\ mission team.
The authors thank Sergei Nayakshin for suggestions and
discussion on ionized skin models, as well as the referee for
useful comments.
This work has made use of data obtained through the High Energy
Astrophysics Science Archive Research Center Online Service, provided by
the NASA Goddard Space Flight Center, the TARTARUS database, which is
supported by Jane Turner and Kirpal Nandra under NASA grants NAG~5-7385
and NAG~5-7067, and the NASA$/$IPAC Extragalactic Database which is
operated by the Jet Propulsion Laboratory, California Institute of
Technology, under contract with the National Aeronautics and Space
Administration. A.M.\ \& R.E.\ acknowledge financial 
support from NASA grant NAG~5-9023.

\appendix
\section{Notes on individual sources}

$\bullet$ 3C~390.3: The short time scale time-resolved and time-average fit 
results are in reasonable agreement with Gliozzi \et\ (2003): the light curve
for $\Gamma$ has absolute values slightly systematically lower by $\lesssim$0.1 than Gliozzi \et\ (2003) but 
the variability patterns are identical. 
On short time scales, a decrease in F$_{2-10}$ by a factor of $\sim$2 over a span of 30~days does 
not appear to have any simultaneous effect on F$_{K\alpha}$, which remains nearly constant during the observation.
On long time scales, there is a doubling in F$_{2-10}$ over 300~days,
and an increase of $\sim$50$\%$ in \fka\ over approximately the same period.

$\bullet$ MCG--6-30-15: Due to improved background models, the current analysis yields
a lower absolute value for $R$ for the integrated spectrum of the short time scale
observation MCG--6-30-15~(I) than in VE01 (values of $R$
obtained from spectral fitting are highly sensitive to
the shape of the modelled background spectrum above $\sim$10~keV),
but the time-resolved spectral analysis and the cross-correlation analysis
on the short time scale observation MCG--6-30-15~(I) is in excellent agreement with VE01.
\fka\ shows variability on short time scales, as well as in the long time scale observation
MCG--6-30-15~(I). The long time scale observation MCG--6-30-15~(II) does not yield a significant detection
of variability in the line; one possibility is that on these time scales probed
($\sim$100~d to $\sim$2000~d) the variability is ``averaged out'' to some degree.
There are examples in these observations of large increases or decreases in F$_{2-10}$ (in some cases by
a factor of $\sim$2) that do not have any obvious immediate effect on F$_{K\alpha}$.
The 6.4~keV dips in the $\fv$ spectra are prominent, especially
for the short time scale observation MCG--6-30-15~(I).
The four observations, when considered in order of increasing duration,
do not show any obvious pattern to the Fe~\ka\ line variability.

$\bullet$ NGC~3516: On short time scales, F$_{2-10}$ and \fka\ exhibit uncorrelated variability.
Persistent trends in F$_{2-10}$ such as the rapid decrease  
from Modified Julian Day (MJD) 50550--50554 are not simultaneously matched by similar behavior in F$_{K\alpha}$.
Rapid changes in \fka\ occur without sudden changes in F$_{2-10}$, e.g., the 
decrease in \fka\ by $\sim$20$\%$ at MJD~50551 and also the increase in \fka\ by $\sim$30$\%$ at 
MJD~50558 as illustrated in Figure~7. The 6.4~keV dip in the short time scale $\fv$ 
spectra is especially noticeable. 
On long time scales, both F$_{2-10}$ and \fka\ 
show an overall decrease by a factor of $\sim$2 over the 1000-day duration.

$\bullet$ NGC~3783: Formally, \fka\ does not demonstrate appreciable variability on either time scale.
Relatively rapid changes in F$_{2-10}$ by $\gtrsim$10$\%$ towards the middle of the short time scale
observation and $\sim$20$\%$~--~$\sim$40$\%$ towards the beginning and end of the
long time scale observation do not appear to have any major simultaneous impact on F$_{K\alpha}$.

$\bullet$ NGC~4151: 
On short time scales, \fka\ is not formally variable. One note of interest, for instance, is the doubling of
F$_{2-10}$ during the last 4 days of the short time scale observation while \fka\ remains virtually constant 
during that time. The 6.4~keV dip in the short time scale $\fv$ spectra is especially noticeable.
On the long time scale, \fka\ is much more variable. There are similar decreasing
trends by a factor of $\sim$3 in both the continuum and line over the whole 800-day duration. 
Of note is the sudden dip and 
rise in the line around MJD~51620 in the long time scale observation while F$_{2-10}$ remains nearly 
constant; Figure~7 demonstrates the dip. 

$\bullet$ NGC~5548: 
On long time scales, a gradual decreasing 
trend by a factor of 3--4 dominates the behavior of both F$_{2-10}$ and F$_{K\alpha}$.
The Fe~\ka\ line variability amplitude is substantially larger on the long time scale than the short
due to this dominating long time scale trend.

$\bullet$ NGC~7469: All the time-average and time-resolved spectral fit
results and cross-correlation results are in reasonable agreement
with Nandra \et\ (2000). The current analysis yields absolute values of $\Gamma$
that are slightly systematically lower by $\lesssim$0.1 and the absolute values of
\fka\ obtained in the current analysis are systematically $\sim$30$\%$ higher compared to Nandra \et\ (2000),
but the variability patterns are consistent.

\clearpage

\begin{deluxetable}{lccccccc}
\tablewidth{6.7in}
\tablenum{1}
\tablecaption{Source and sampling parameters\label{tab1}}
\tablehead{
\colhead{Source} & \colhead{2--10~keV Lum.} & \colhead{} & \colhead{MJD~Date} & \colhead{Mean} & \colhead{Spectral Fits}  & \colhead{Spectral Fits}    \\
\colhead{Name} & \colhead{(log(erg s$^{-1}$))} & \colhead{z} & \colhead{Range} & \colhead{c s$^{-1}$} & \colhead{Time Bin} & \colhead{No.\ Pts}   }
\startdata
3C~390.3   &      44.26  & 0.056  &  50220.6--50278.0  &   2.3  &  6.0 d  &   10 \\
MCG--6-30-15 I  & 42.80  & 0.008  &  50664.1--50672.5  &   4.1  &  0.13 d &  63 \\
MCG--6-30-15 II & 42.83  & 0.008  &  51622.7--51676.0  &   4.4  &  1.6 d  &   34 \\
NGC~3516        & 42.90  & 0.009  &  50523.0--50657.4  &   4.2  &  6.4 d  &   21 \\
NGC~3783        & 43.12  & 0.010  &  51960.1--51980.0  &   5.7  &  1.7 d  &   12 \\
NGC~4151        & 42.24  & 0.003  &  51870.6--51904.9  &   7.4  &  1.4 d  &   24 \\
NGC~5548        & 43.52  & 0.017  &  52091.6--52126.7  &   4.5  &  3.2 d  &   11 \\
NGC~7469        & 43.25  & 0.016  &  50244.0--50275.7  &   2.8  &  1.3 d  &   25 \\ \hline
3C~390.3      &   44.40 & 0.056 & 51186.0--51964.8  &  3.2  & 51.9 d &   15  \\
MCG--6-30-15 I  & 42.81 & 0.008 & 51926.7--52252.9  &  4.2  & 23.4 d &   13 \\ 
MCG--6-30-15 II & 42.85 & 0.008 & 50159.8--52252.9  &  4.6  & 128.7 d &  16 \\ 
NGC~3516        & 42.86 & 0.009 & 50523.0--51593.4  &  3.8   & 53.5 d &  20 \\ 
NGC~3783        & 43.22 & 0.010 & 51180.5--52375.2  &  7.1   & 59.8 d &  20 \\ 
NGC~4151        & 42.55 & 0.003 & 51179.6--51964.6  &  14.9  & 43.6 d &  18 \\ 
NGC~5548        & 43.55 & 0.017 & 51183.9--52643.1  &  4.8   & 104.2 d & 14 \\ 
\enddata
\tablecomments{The short time scale data are listed in the top half of the table; 
the bottom half lists the long time scale data.
Column (2), log of the 2-10~keV luminosity, was calculated using the online
W3PIMMS tool and using the mean 
2--10~keV count rate per PCU, X-ray photon indices and neutral absorption
based on the time-averaged spectral fitting below, and 
assuming $H_{0}=75$~km~s$^{-1}$~Mpc$^{-1}$
and $q_{0} =0.5 $. The count rates in Column (5) are 2--10~keV count rates
per PCU.}
\end{deluxetable}

\begin{deluxetable}{lccccccc}
\tablewidth{6.6in}
\tablenum{2}
\tablecaption{Time-average fit parameters, single-Gaussian model  \label{tab2}}
\tablehead{
\colhead{Source} & \colhead{} & \colhead{} & \colhead{$E_o$} & \colhead{$\sigma$} & \colhead{} & \colhead{} & \colhead{} \\ 
\colhead{Name} & \colhead{N$_{{\rm H}}$ } & \colhead{$\Gamma$} & \colhead{(keV)} & \colhead{(keV)} & \colhead{$F_{K\alpha}$} & \colhead{$R$} & \colhead{$\chi^2$/d.o.f.} }
\startdata
3C~390.3         & 4.28  &  1.641$\pm$0.006           & 6.00$^{+0.04}_{-0.06}$ & 0.44$\pm$0.07          & 0.64$\pm$0.06          & $<$ 0.03 &   61.0/34\\
MCG--6-30-15~I   & 4.09  &  2.018$\pm$0.010           & 5.87$\pm$0.02          & 0.68$\pm$0.02          & 1.79$\pm$0.06         & 0.85$\pm$0.07    &    292.0/33\\
MCG--6-30-15 II  & 4.09  &  1.947$\pm$0.016           & 6.17$^{+0.02}_{-0.03}$ & 0.46$\pm$0.04          & 1.35$\pm$0.07         & 1.15$\pm$0.12    &    23.0/27  \\  
NGC~3516         & 2.94  &  1.573$\pm$0.007           & 6.02$\pm$0.01          & 0.59$^{+0.01}_{-0.02}$ & 2.62$^{+0.06}_{-0.04}$ & 0.20$^{+0.04}_{-0.07}$    &   487.6/33\\
NGC~3783         & 8.26  &  1.666$^{+0.022}_{-0.019}$ & 6.27$^{+0.04}_{-0.03}$ & 0.41$\pm$0.06          & 1.75$\pm$0.12         & 0.62$\pm$0.12    &   22.3/28\\
NGC~4151         &  616  &  1.524$\pm$0.010           & 6.40$\pm$0.01          & 0.14$^{+0.04}_{-0.03}$ & 3.45$\pm$0.08         & 0.55$\pm$0.05     &  101.0/26\\
NGC~5548         & 1.75  &  1.733$\pm$0.017           & 6.27$\pm$0.06          & 0.42$\pm$0.10          & 0.73$^{+0.07}_{-0.08}$  & 0.44$\pm$0.10     &  25.3/28 \\
NGC~7469         & 4.86  &  1.845$\pm$0.009           & 6.31$\pm$0.02          & 0.28$\pm$0.04          & 0.64$\pm$0.02          & 0.42$\pm$0.05     &  146.0/33\\ \hline
3C~390.3       &   4.28  & 1.646$^{+0.012}_{-0.008}$  & 6.10$\pm$0.04          & 0.25$^{+0.08}_{-0.11}$ & 0.47$\pm$0.04          & 0.03$^{+0.06}_{-0.02}$  &  122.0/97    \\
MCG--6-30-15~I  &  4.09  & 2.020$^{+0.030}_{-0.029}$  & 5.92$^{+0.07}_{-0.08}$ & 0.75$^{+0.10}_{-0.09}$ & 1.76$\pm$0.02         & 1.96$^{+0.28}_{-0.25}$  &  30.75/28    \\
MCG--6-30-15 II &  4.09  & 2.000$\pm$0.006            & 6.00$\pm$0.01          & 0.59$^{+0.01}_{-0.02}$ & 1.75$\pm$0.03         & 1.17$^{+0.05}_{-0.04}$  &  726.9/98    \\
NGC~3516       &   2.94  & 1.642$^{+0.005}_{-0.004}$  & 6.09$\pm$0.01          & 0.47$^{+0.02}_{-0.01}$ & 2.11$\pm$0.03         & 0.27$\pm$0.03  & 1009.7/65    \\
NGC~3783       &   8.26  & 1.693$^{+0.008}_{-0.007}$  & 6.23$\pm$0.02          & 0.43$\pm$0.03          & 2.04$\pm$0.06         & 0.47$^{+0.05}_{-0.04}$  & 206.6/98    \\
NGC~4151       &   812   & 1.692$^{+0.014}_{-0.005}$  & 6.40$^{+0.05}_{-0.01}$ & 0.12$^{+0.02}_{-0.12}$ & 3.72$^{+0.04}_{-0.08}$ & 0.55$^{+0.04}_{-0.01}$  &  581.6/94    \\
NGC~5548       &   1.75  & 1.739$\pm$0.010            & 6.28$^{+0.02}_{-0.03}$ & 0.36$^{+0.05}_{-0.06}$ & 0.76$^{+0.04}_{-0.05}$  & 0.41$^{+0.02}_{-0.08}$  &  105.0/99    \\
\enddata
\tablecomments{Results of fits of the model (power-law + {\sc pexrav} + Gaussian)$\times$absorption to 
the time-average data for all targets. The short time scale data are listed in the top half of the table; 
the bottom half lists the long time scale data.
The {\sc pexrav} component was omitted for the short time scale spectrum of 3C~390.3.
Column (2) lists the neutral absorption column in units of 10$^{20}$~cm$^{-2}$,
modelled using the Galactic column for all sources except NGC~4151.
The line energy, $E_o$, Column (4), is the energy in the observed, not the rest, frame.
Column (5) lists $\sigma$, the FWHM of the Gaussian.
Column (6) lists the Fe~\ka\ line normalization in units of 10$^{-4}$ ph cm$^{-2}$ s$^{-1}$.
Column (7) lists $R$, the relative reflection 
of the Compton reflection hump.
Errors listed above are 1$\sigma$ errors determined by XSPEC.}
\end{deluxetable}

\begin{deluxetable}{lcccccccc}
\footnotesize
\tablewidth{6.9in}
\tablenum{3}
\tablecaption{Time-average fit parameters, Laor model \label{tab3}}
\tablehead{
\colhead{Source} & \colhead{} & \colhead{} & \colhead{$E_o$} & \colhead{} & \colhead{} & \colhead{} & \colhead{} & \colhead{} \\ 
\colhead{Name} & \colhead{N$_{{\rm H}}$} & \colhead{$\Gamma$} & \colhead{(keV)} & \colhead{$\beta$} & \colhead{$R_{in}$} & \colhead{$F_{K\alpha}$} & \colhead{$R$} & \colhead{$\chi^2$/d.o.f.} }
\startdata
3C~390.3        & 4.28 & 1.637$^{+0.007}_{-0.005}$ & 6.19$\pm$0.5           & 3.8$^{+6.1}_{-1.0}$ & 13$^{+7}_{-6}$       & 0.68$^{+0.06}_{-0.03}$  & $<$ 0.02               & 55.6/33\\
MCG--6-30-15~I  & 4.09 & 1.922$^{+0.019}_{-0.015}$ & 6.47$^{+0.02}_{-0.03}$ & 3.2$\pm$0.1         & 2.2$^{+0.8}_{-1.0}$  & 2.76$^{+0.22}_{-0.26}$ & 0.45$\pm$0.08          & 98.2/32\\
MCG--6-30-15 II & 4.09 & 1.932$^{+0.018}_{-0.016}$ & 6.34$\pm$0.3           & 3.6$^{+2.1}_{-0.7}$ & 13$^{+6}_{-4}$       & 1.37$\pm$0.07         & 1.06$\pm$0.12          & 19.6/25\\
NGC~3516        & 2.94 & 1.507$\pm$0.004           & 6.48$^{+0.01}_{-0.03}$ & 8.0$^{+1.9}_{-1.1}$ & 7.1$^{+1.1}_{-0.1}$  & 3.09$\pm$0.04         & $<$ 0.03               & 101.5/32\\
NGC~3783        & 8.26 & 1.650$\pm$0.022           & 6.46$\pm$0.5           & 2.5$^{+0.1}_{-0.4}$ & 5.8$^{+4.8}_{-4.2}$  & 1.95$^{+0.27}_{-0.37}$ & 0.57$^{+0.06}_{-0.19}$ & 20.4/27   \\    
NGC~4151        & 618  & 1.524$^{+0.011}_{-0.008}$ & 6.42$\pm$0.1           & 3.0$^{+6.9}_{-3.0}$ & 300$^{+100}_{-275}$  & 3.43$^{+0.10}_{-0.06}$ & 0.55$\pm$0.05          & 101.0/25\\
NGC~5548        & 1.75 & 1.725$^{+0.017}_{-0.025}$ & 6.43$^{+0.15}_{-0.07}$ & 2.8$^{+7.1}_{-0.8}$ & 8.7$^{+24}_{-7.5}$ & 0.78$^{+0.32}_{-0.11}$  & 0.40$^{+0.10}_{-0.09}$ & 24.5/27\\
NGC~7469        & 4.86 & 1.832$\pm$0.008           & 6.43$\pm$0.2           & 2.2$\pm$0.1         & 2.8$\pm$1.4          & 0.79$^{+0.08}_{-0.03}$  & 0.38$^{+0.03}_{-0.06}$ & 117.4/32\\ \hline
3C~390.3        & 4.28 & 1.653$^{+0.001}_{-0.002}$ & 6.12$\pm$0.4           & 8.2$^{+1.7}_{-8.1}$ & 260$^{+140}_{-200}$  & 0.44$^{+0.02}_{-0.03}$  & 0.06$^{+0.06}_{-0.01}$ & 122.1/96 \\
MCG--6-30-15~I  & 4.09 & 1.975$^{+0.021}_{-0.030}$ & 6.41$\pm$0.9           & 9.9$^{+0.0}_{-6.9}$ & 8.9$^{+1.0}_{-2.3}$  & 1.70$^{+0.15}_{-0.10}$ & 1.59$^{+0.22}_{-0.07}$ & 27.1/27  \\
MCG--6-30-15 II & 4.09 & 1.919$^{+0.005}_{-0.010}$ & 6.43$^{+0.01}_{-0.02}$ & 2.9$\pm$0.1         & 1.8$^{+0.4}_{-0.5}$  & 2.64$^{+0.01}_{-0.07}$ & 0.81$^{+0.03}_{-0.05}$ & 457.0/97  \\
NGC~3516        & 2.94 & 1.582$\pm$0.002           & 6.37$\pm$0.1           & 2.6$\pm$0.1         & 2.8$^{+0.1}_{-0.2}$  & 2.73$^{+0.02}_{-0.09}$ & 0.08$\pm$0.02          & 766.8/64  \\
NGC~3783        & 8.26 & 1.664$^{+0.003}_{-0.009}$ & 6.43$^{+0.03}_{-0.02}$ & 2.9$^{+0.2}_{-0.1}$ & 6.3$\pm$0.8          & 2.37$^{+0.17}_{-0.03}$ & 0.38$\pm$0.03          & 185.4/97  \\
NGC~4151        & 690  & 1.604$^{+0.010}_{-0.011}$ & 6.39$\pm$0.1           & 0.0$^{+9.9}_{-0.0}$ & 230$^{+170}_{-60}$   & 4.04$^{+0.1}_{-0.2}$ & 0.54$^{+0.02}_{-0.03}$ & 450.5/91   \\
NGC~5548        & 1.75 & 1.732$\pm$0.011           & 6.38$\pm$0.3           & 8.6$^{+1.3}_{-5.3}$ & 44$^{+1}_{-25}$      & 0.73$^{+0.04}_{-0.03}$  & 0.40$^{+0.05}_{-0.03}$ & 120.5/97   \\
\enddata
\tablecomments{Fit parameters for the Laor model fits to the time-averaged data.
The short time scale data are listed in the top half of the table; 
the bottom half lists the long time scale data.
The {\sc pexrav} component was omitted for
the short time scale spectra of 3C~390.3 and NGC~3516.
Column (2) lists the neutral absorption column in units of 10$^{20}$~cm$^{-2}$,
modelled using the Galactic column for all sources except NGC~4151.
Column~(4) is the emissivity index $\beta$. 
Column~(5) is $R_{in}$, the innermost disk radius of line 
emission, in units of GM/c$^2$ where M is the black hole mass.
Column (7) lists the Fe~\ka\ line normalization in units of 10$^{-4}$ ph cm$^{-2}$ s$^{-1}$.
Column (8) lists $R$, the relative reflection 
of the Compton reflection hump.
Errors listed above are 1$\sigma$ errors determined by XSPEC.}
\end{deluxetable}

\begin{deluxetable}{lcccccc}
\tablewidth{4.7in}
\tablenum{4}
\tablecaption{F-test results to thaw $\Gamma$, F$_{K\alpha}$ or $R$ in the single-Gaussian model \label{tab4}}
\tablehead{
\colhead{Source} & \colhead{$\Gamma$} & \colhead{$\Gamma$} & \colhead{F$_{K\alpha}$}  & \colhead{F$_{K\alpha}$} & \colhead{$R$} & \colhead{$R$}  \\
\colhead{Name} & \colhead{F} & \colhead{Prob} & \colhead{F} & \colhead{Prob} & \colhead{F} & \colhead{Prob}  }         
\startdata
3C~390.3        &  5.8  & 1.2E-10 &   1.80 &  3.7 E-2 &  N/A   &  N/A      \\    
MCG--6-30-15 I  &  1.78 & 1.9E-4  &   1.63 &  1.4E-3  &  0.76  &  0.92      \\ 
MCG--6-30-15 II &  2.4  & 1.6E-5  &   1.70 &  5.7E-3  &  1.31  &  0.11      \\
NGC~3516        &  24.5 & $<$1$\times$10$^{-40}$       &   3.20 &  2.9E-6  &  0.39  &  0.99     \\
NGC~3783        &  0.81 & 0.64    &   0.72 &  0.74    &  0.53  &  0.90     \\
NGC~4151        &  1.94 & 4.8E-3  &   0.82 &  0.71    &  1.04  &  0.41     \\
NGC~5548        &  1.5  & 0.26    &   0.99 &  0.45    &  0.47  &  0.92     \\
NGC~7469        &  3.80 & 1.7E-9  &   2.05 &  1.9E-3  &  1.58  &  3.5E-2      \\  \hline
3C~390.3         & 5.84   & 3.2E-11 &  0.94  & 0.52    &  0.65 &  0.84   \\
MCG--6-30-15 I   & 2.21   & 2.9E-2  &  1.65  & 4.4E-2  &  1.80 &  2.3E-2 \\    
MCG--6-30-15~II  & 2.90   & 1.3E-4  &  2.29  & 2.9E-3  &  0.86 &  0.62   \\   
NGC~3516         & 13.0   & 2.3E-36 &  4.83  & 3.8E-11 &  1.08 &  0.36   \\
NGC~3783         & 1.17   & 0.28    &  1.03  & 0.43    &  0.53 &  0.95   \\
NGC~4151         & 12.3   & 5.2E-31 &  7.74  & 2.2E-18 &  1.76 &  2.7E-2 \\
NGC~5548         & 1.77   & 4.11E-2 &  1.83  & 3.2E-2  &  0.85 &  0.61  \\
\enddata 
\tablecomments{Results of F-test to determine which parameters to thaw in the time-resolved fits.
High values of the F-statistic and low values of the probability (of observing that
value of F from a random set of data)
indicate that the fits show significant improvement to thaw that parameter.
The short time scale data are listed in the top half of the table; 
the bottom half lists the long time scale data.}
\end{deluxetable}

\begin{deluxetable}{lccc}
\tablewidth{5.2in}
\tablenum{5}
\tablecaption{Mean spectral fit variability parameters and errors\label{tab5}}
\tablehead{
\colhead{Source} & \colhead{Mean $F_{2-10}$} & \colhead{Mean} & \colhead{Mean F$_{K\alpha}$} \\
\colhead{Name}  &  \colhead{(10$^{-11}$ ph cm$^{-2}$ s$^{-1}$)} & \colhead{$\Gamma$} & \colhead{(10$^{-4}$ ph cm$^{-2}$ s$^{-1}$)}  }
\startdata
3C~390.3         & 2.58 $\pm$ 0.05 &  1.633 $\pm$ 0.023 & 0.639 $\pm$  0.174    \\
MCG--6-30-15 I   & 4.73 $\pm$ 0.05 &  2.005 $\pm$ 0.063 & 1.829 $\pm$ 0.250   \\ 
MCG--6-30-15 II  & 4.58 $\pm$ 0.10 &  1.914 $\pm$ 0.103 & 1.384 $\pm$ 0.261     \\ 
NGC~3516         & 4.80 $\pm$ 0.08 &  1.609 $\pm$ 0.061 & 2.384 $\pm$ 0.177     \\ 
NGC~3783         & 6.15 $\pm$ 0.14 &  1.665 $\pm$ 0.028 & 1.768 $\pm$ 0.200     \\ 
NGC~4151         & 10.11 $\pm$ 0.19 &  1.519 $\pm$ 0.037 & 3.466 $\pm$ 0.264     \\ 
NGC~5548         & 4.85 $\pm$ 0.13 &  1.723 $\pm$ 0.028 & 0.724 $\pm$ 0.149      \\ 
NGC~7469         & 3.24 $\pm$ 0.05 &  1.845 $\pm$ 0.034 & 0.642 $\pm$ 0.096      \\  \hline
3C~390.3         & 3.44 $\pm$ 0.08 &  1.644 $\pm$ 0.044 & 0.453 $\pm$  0.066     \\
MCG--6-30-15 I   & 5.10 $\pm$ 0.14 &  2.002 $\pm$ 0.080 & 1.788 $\pm$ 0.253    \\
MCG--6-30-15 II  & 5.41 $\pm$ 0.10 &  2.014 $\pm$ 0.038 & 1.801 $\pm$ 0.226  \\
NGC~3516         & 4.31 $\pm$ 0.08 &  1.583 $\pm$ 0.061 & 1.953 $\pm$ 0.269  \\
NGC~3783         & 7.56 $\pm$ 0.12 &  1.689 $\pm$ 0.037 & 2.080 $\pm$ 0.202   \\
NGC~4151         & 22.23 $\pm$ 0.24 &  1.616 $\pm$ 0.055 & 4.878 $\pm$ 0.795  \\
NGC~5548         & 4.12 $\pm$ 0.10 &  1.711 $\pm$ 0.041 & 0.750 $\pm$ 0.104   \\
\enddata
\tablecomments{The short time scale data are listed 
in the top half of the table; 
the bottom half lists the long time scale data.
Listed are the mean values of $F_{2-10}$, $\Gamma$ and $F_{K\alpha}$.
Errors for $\Gamma$ and \fka\ were derived using the
formulation of $\S$3.3.
The $F_{2-10}$ errors listed above
are the average of the errors for $F_{2-10}$ in a given time bin,
which were derived 
from the 2--10~keV continuum light curve,
using the mean flux error 
on the $\sim$1~ks exposures in that time bin.}
\end{deluxetable}

\begin{deluxetable}{lcccc}
\tablewidth{3.5in}
\tablenum{6}
\tablecaption{$\fv$ values \label{tab6}}
\tablehead{
\colhead{Source} & \colhead{Time} & \colhead{Continuum} & \colhead{Line} \\
\colhead{Name}   & \colhead{Scale} & \colhead{$\fv$ ($\%$)} & \colhead{$\fv$ ($\%$)} }
\startdata
3C~390.3      &   Short     &     29.3 $\pm$ 0.6   &   Undef.        \\
              &   Long      &     26.9 $\pm$ 0.6   &   13.7 $\pm$ 4.7 \\
MCG--6-30-15 &    Short I   &    18.5 $\pm$ 0.1    &   9.4 $\pm$  2.5   \\ 
             &    Short II  &    17.4 $\pm$ 0.4    &   14.4 $\pm$  4.4   \\ 
             &    Long I    &    16.1 $\pm$ 0.7    &   14.1 $\pm$  4.8   \\    
             &    Long II   &    7.0 $\pm$ 0.5     &   Undef. \\               
NGC~3516     &    Short     &    24.6 $\pm$ 0.4    &   5.6 $\pm$  2.2   \\ 
             &    Long      &    32.9 $\pm$ 0.4    &   14.9 $\pm$  3.7 \\   
NGC~3783     &    Short     &     4.7 $\pm$ 0.7           &   6.0 $\pm$ 5.5 \\
             &    Long      &     10.0 $\pm$ 0.4   &   1.5 $\pm$ 9.9 \\
NGC~4151     &    Short      &    16.7 $\pm$ 0.4   &   3.2  $\pm$ 3.1 \\
             &    Long       &    30.5 $\pm$ 0.2   &   24.0 $\pm$ 4.3 \\
NGC~5548     &    Short       &    24.8 $\pm$ 0.8   &   8.2 $\pm$ 12.6 \\
             &    Long        &   39.0 $\pm$ 0.7   &   28.8 $\pm$ 3.9 \\
NGC~7469     &    Short       &   13.3 $\pm$ 0.3   &   Undef. \\ 
\enddata 
\tablecomments{Undefined fractional variability measurements indicate
a measured variance that is smaller than that expected solely from
measurement noise.}  
\end{deluxetable}

\begin{deluxetable}{lccccc}
\tablewidth{4.4in}
\tablenum{7}
\tablecaption{F$_{2-10}$--F$_{K\alpha}$ Cross-Correlation Results \label{tab7}}
\tablehead{
\colhead{Source} & \colhead{Time} & \colhead{DCF} & \colhead{DCF} & \colhead{ICF} & \colhead{ICF} \\
\colhead{Name}   & \colhead{Scale} & \colhead{$\tau$(d)} &  \colhead{r$_{max}$} & \colhead{$\tau$(d)} &  \colhead{r$_{max}$} }
\startdata
3C~390.3 & Short    & +12.0 & 0.46    & +9$\pm$20.2 & 0.61\\
MCG--6-30-15 I &    & +1.3  & 0.51    & +1.4$\pm$1.5 & 0.53 \\
MCG--6-30-15 II &   & --14.4& 0.38    & --14.4$\pm$18.0 & 0.38 \\
NGC~3516        &   & +6.4  & 0.17    & +6.4$\pm$41.2 & 0.18 \\
NGC~3783        &   & +3.3  & 0.59    & +4.1$\pm$5.4 & 0.72 \\
NGC~4151        &   & +2.8  & 0.23    & +3.5$\pm$10.0 & 0.27 \\
NGC~5548        &   & +6.4  & 0.63    & +8.0$\pm$6.1 & 0.68 \\
NGC~7469        &   & 0 & 0.57        & +0.6$\pm$7.2 & 0.67 \\ \hline
3C~390.3 & Long     & +51.9 & 0.48    & +25.9$\pm$159 & 0.48 \\
MCG--6-30-15 I &    & 0 & 0.75        & 0$\pm$65.6 & 0.73 \\  
MCG--6-30-15 II &   & --386 & 0.45    & 0$\pm$606 & 0.45  \\  
NGC~3516        &   & --161 & 0.66    & --134$\pm$262 & 0.71 \\
NGC~3783        &   & 0 & 0.53        & 0$\pm$388 & 0.51  \\
NGC~4151        &   & 0 & 0.56    & 0$\pm$258 & 0.61 \\
NGC~5548        &   & 0 & 0.84     & 0$\pm$428 & 0.87  \\ 
\enddata 
\tablecomments{Lags are defined such that a positive lag means the continuum light curve leads the line light curve.}
\end{deluxetable}

\begin{deluxetable}{lccccc}
\tablewidth{4.4in}
\tablenum{8}
\tablecaption{F$_{2-10}$--$\Gamma$ Cross-Correlation Results \label{tab8}}
\tablehead{
\colhead{Source} & \colhead{Time} & \colhead{DCF} & \colhead{DCF} & \colhead{ICF} & \colhead{ICF} \\
\colhead{Name}   & \colhead{Scale} & \colhead{$\tau$(d)} &  \colhead{r$_{max}$} & \colhead{$\tau$(d)} &  \colhead{r$_{max}$} }
\startdata
3C~390.3 & Short    & 0 & 0.91      & --3.0$\pm$7.8 & 0.92 \\
MCG--6-30-15 I &    & 0 & 0.82      & 0$\pm$1.4 & 0.82 \\
MCG--6-30-15 II &   & 0 & 0.54      & --0.8$\pm$10.9 & 0.57 \\
NGC~3516        &   & 0 & 0.90      & 0$\pm$10.6 & 0.90 \\
NGC~3783        &   & 0 & 0.88      & 0$\pm$5.7 & 0.89 \\
NGC~4151        &   & 0 & 0.69      & 0$\pm$8.4 & 0.73 \\
NGC~5548        &   & --9.6 & 0.85  & --8.0$\pm$5.3 & 0.88 \\
NGC~7469        &   & --3.8 & 0.64  & --4.4$\pm$8.4 & 0.73 \\  \hline
3C~390.3 & Long     & 0 & 0.85      & 0$\pm$136 & 0.82 \\
MCG--6-30-15 I &    & 0 & 0.94      & 0$\pm$66.6 & 0.86 \\
MCG--6-30-15 II &   & 0 & 0.62      & 0$\pm$555 & 0.64 \\
NGC~3516        &   & 0 & 0.92      & 0$\pm$161 & 0.93 \\
NGC~3783        &   & 0 & 0.81      & 0$\pm$357 & 0.81 \\
NGC~4151        &   & 0 & 0.31      & +21.8$\pm$191 & 0.46 \\ 
NGC~5548        &   & 0 & 0.74      & --52.1$\pm$506 & 0.75 \\
\enddata 
\tablecomments{Lags are defined such that a positive lag means the continuum light curve leads the $\Gamma$ light curve.}
\end{deluxetable}

\begin{deluxetable}{lccccc}
\tablewidth{4.4in}
\tablenum{9}
\tablecaption{$\Gamma$--F$_{K\alpha}$ Cross-Correlation Results \label{tab9}}
\tablehead{
\colhead{Source} & \colhead{Time} & \colhead{DCF} & \colhead{DCF} & \colhead{ICF} & \colhead{ICF} \\
\colhead{Name}   & \colhead{Scale} & \colhead{$\tau$(d)} &  \colhead{r$_{max}$} & \colhead{$\tau$(d)} &  \colhead{r$_{max}$} }
\startdata
3C~390.3 & Short    & +12.0 & 0.64      & +9.0$\pm$18.1 & 0.66\\
MCG--6-30-15 I &    & +1.5 & 0.49      &  +1.4$\pm$2.0 & 0.54 \\
MCG--6-30-15 II &   & +9.6 & 0.38       &  +10.4$\pm$18.2 & 0.44 \\
NGC~3516        &   & --32.0 & 0.25     &  --35.2$\pm$42.9 & 0.33 \\ 
NGC~3783        &   & +3.3  & 0.70      &  +4.1$\pm$5.6 & 0.74  \\
NGC~4151        &   & +4.2 & 0.28      &  +4.9$\pm$11.6 & 0.33 \\
NGC~5548        &   & +9.6 & 0.43       &  +9.6$\pm$11.3 & 0.47 \\
NGC~7469        &   & +5.7 & 0.40       &  +5.7$\pm$9.6 & 0.45 \\ \hline
3C~390.3 & Long     & +51.9 & 0.52      &  +25.9$\pm$156 & 0.56 \\
MCG--6-30-15 I &    & 0 & 0.67          & --11.7$\pm$97.0 & 0.63 \\
MCG--6-30-15 II &   & --386 & 0.47      & --386$\pm$667 & 0.44 \\
NGC~3516        &   & --107 & 0.79      & --134$\pm$300 & 0.82 \\
NGC~3783        &   & 0 & 0.47          & 0$\pm$390 & 0.46 \\
NGC~4151        &   & --43.6 & 0.19     & --174$\pm$280 & 0.17 \\
NGC~5548        &   & 0 & 0.55       & +52.1$\pm$481 & 0.61 \\ 
\enddata 
\tablecomments{Lags are defined such that a positive lag means the first light curve leads the second.}
\end{deluxetable}

\begin{deluxetable}{lccccc}
\tablewidth{3.7in}
\tablenum{10}
\tablecaption{$\fv$ spectral sampling parameters \label{tab10}}
\tablehead{
\colhead{Source} & \colhead{MJD~Date} & \colhead{Time Bin} & \colhead{No.} \\
\colhead{Name}   & \colhead{Range}    & \colhead{Size (d)} & \colhead{Pts} }
\startdata
3C~120          & 50458.5--50515.3  &   2.0  & 29\\
3C~390.3        & 50220.6--50278.0  &   2.0  & 29\\
Akn~564         & 51694.8--51726.7  &  1.6 & 20\\
Fairall~9       & 52144.9--52179.0  &  1.6 & 21\\
IC~4329a        &  50665.8--50723.5  & 2.0 & 28 \\
MCG--6-30-15 I  & 50664.1--50672.5  & 0.13 & 63\\
MCG--6-30-15 II &  51622.7--51676.0  & 1.6 & 34\\
NGC~3516        & 50523.0--50657.4  &  4.8 & 28\\
NGC~3783        &  51960.1--51980.0  & 0.66 & 30\\
NGC~4151        &  51870.6--51904.9  &  1.2 & 29\\
NGC~5548        & 52091.6--52126.7  & 1.3 &  27\\
NGC~7469        &  50244.0--50275.7  & 1.1 &  30\\   \hline
3C~120          & 50812.4--51259.7 & 20.4 & 19 \\
3C~390.3        & 51259.7--51676.0 & 20.8 & 20 \\
Akn~120         & 50868.1--51259.7 & 16.3 & 19 \\
Akn~564         & 51678.7--52683.7 & 50.2 & 20 \\
Fairall~9       & 51679.8--52678.2 & 50.0 & 20 \\
MCG--6-30-15 I  & 51926.7--52252.9 & 13.1 & 23 \\
MCG--6-30-15 II & 50188.9--51259.7 & 53.6 & 20 \\
NGC~3516        & 50523.0--51259.7 & 28.3 & 26 \\
NGC~3783        & 51679.7--52375.9 & 29.0 & 24 \\
NGC~4051        & 51676.0--52228.3 & 27.6 & 20 \\
NGC~4151        & 51259.7--51676.0 & 17.3 & 24 \\
NGC~5548        & 51678.7--52681.4 & 45.6 & 22 \\
\enddata
\end{deluxetable}


\clearpage 

\begin{figure}
\figurenum{1}
\epsscale{0.8}
\plotone{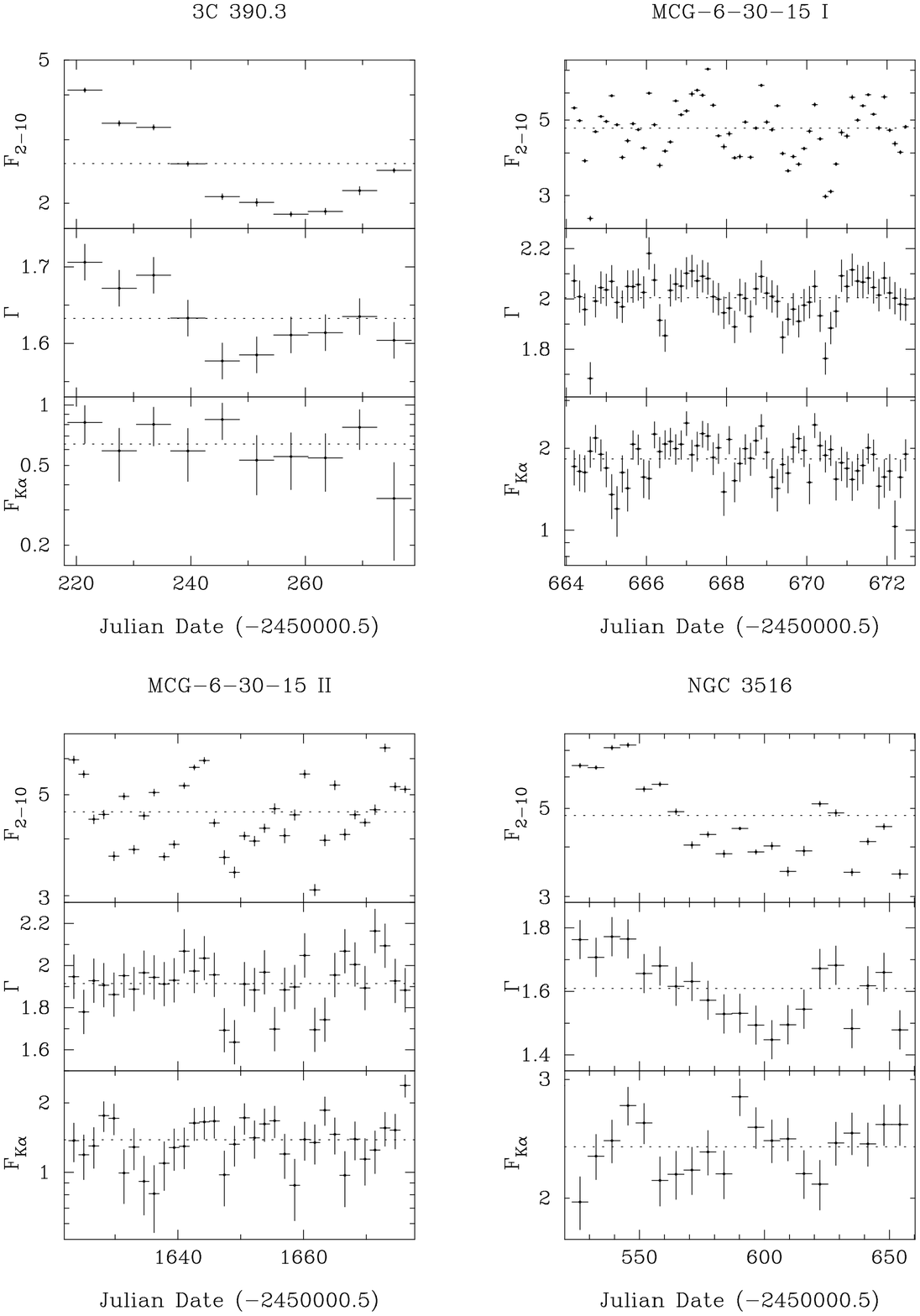}
\caption{Short time scale light curves for continuum flux, $\Gamma$ and Fe~\ka\ line flux. The continuum flux is
in units of 10$^{-11}$ photons cm$^{-2}$ s$^{-1}$. 
The Fe~\ka\ flux is
in units of 10$^{-4}$ photons cm$^{-2}$ s$^{-1}$. 
The dotted lines respresent the mean values of each parameter.}
\end{figure}
\clearpage 
\setcounter{figure}{0}
\begin{figure}
\figurenum{1}
\epsscale{0.8}
\plotone{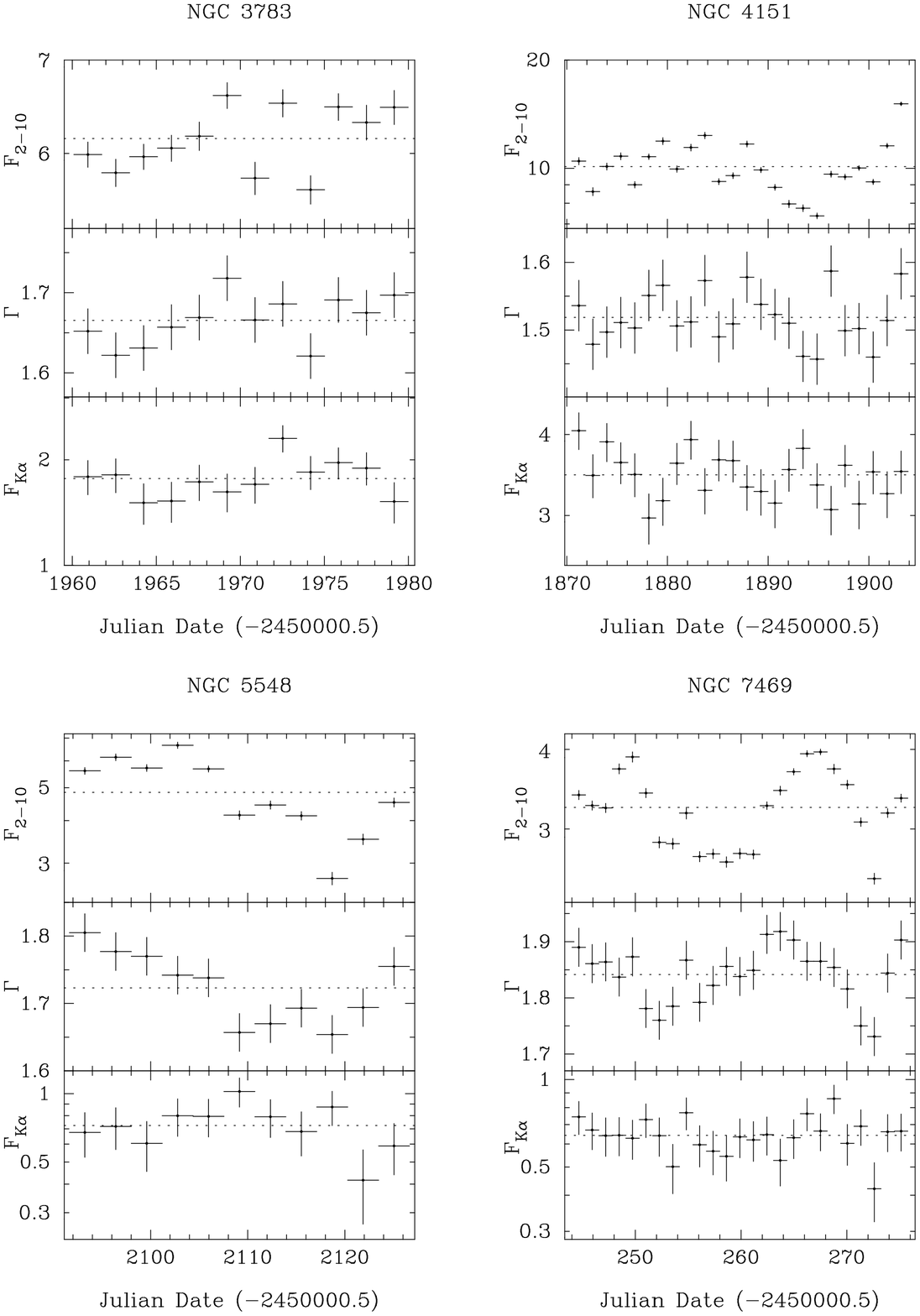}
\caption{Figure 1, cont'd.}
\end{figure}
\clearpage

\begin{figure}
\figurenum{2}
\epsscale{0.8}
\plotone{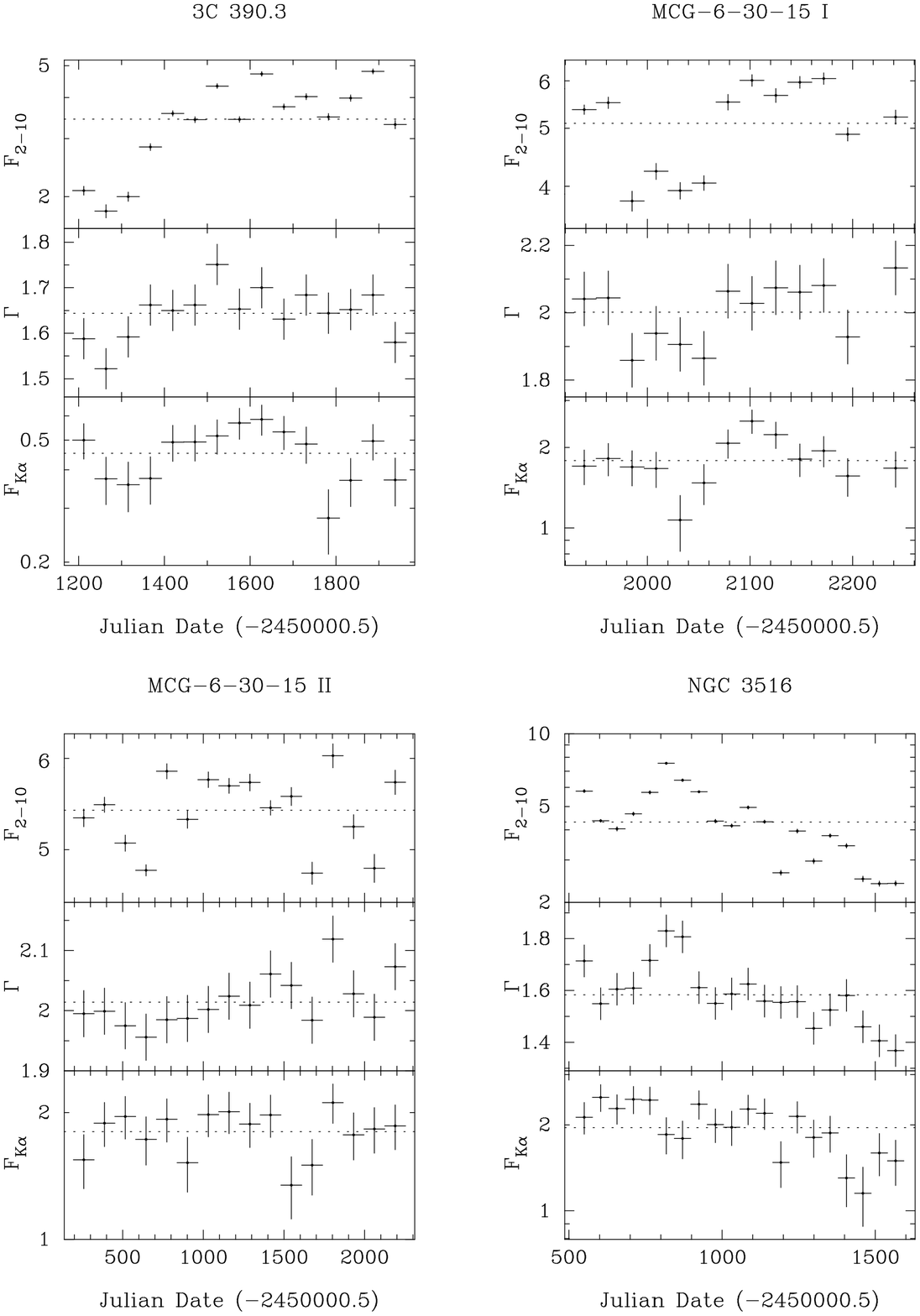}
\caption{Long time scale light curves for 
continuum flux, $\Gamma$ and Fe~\ka\ line flux. The continuum flux is
in units of 10$^{-11}$ photons cm$^{-2}$ s$^{-1}$. 
The Fe~\ka\ flux is
in units of 10$^{-4}$ photons cm$^{-2}$ s$^{-1}$. 
The dotted lines respresent the mean values of each parameter.}
\end{figure}
\clearpage 
\setcounter{figure}{1}
\begin{figure}
\figurenum{2}
\epsscale{0.8}
\plotone{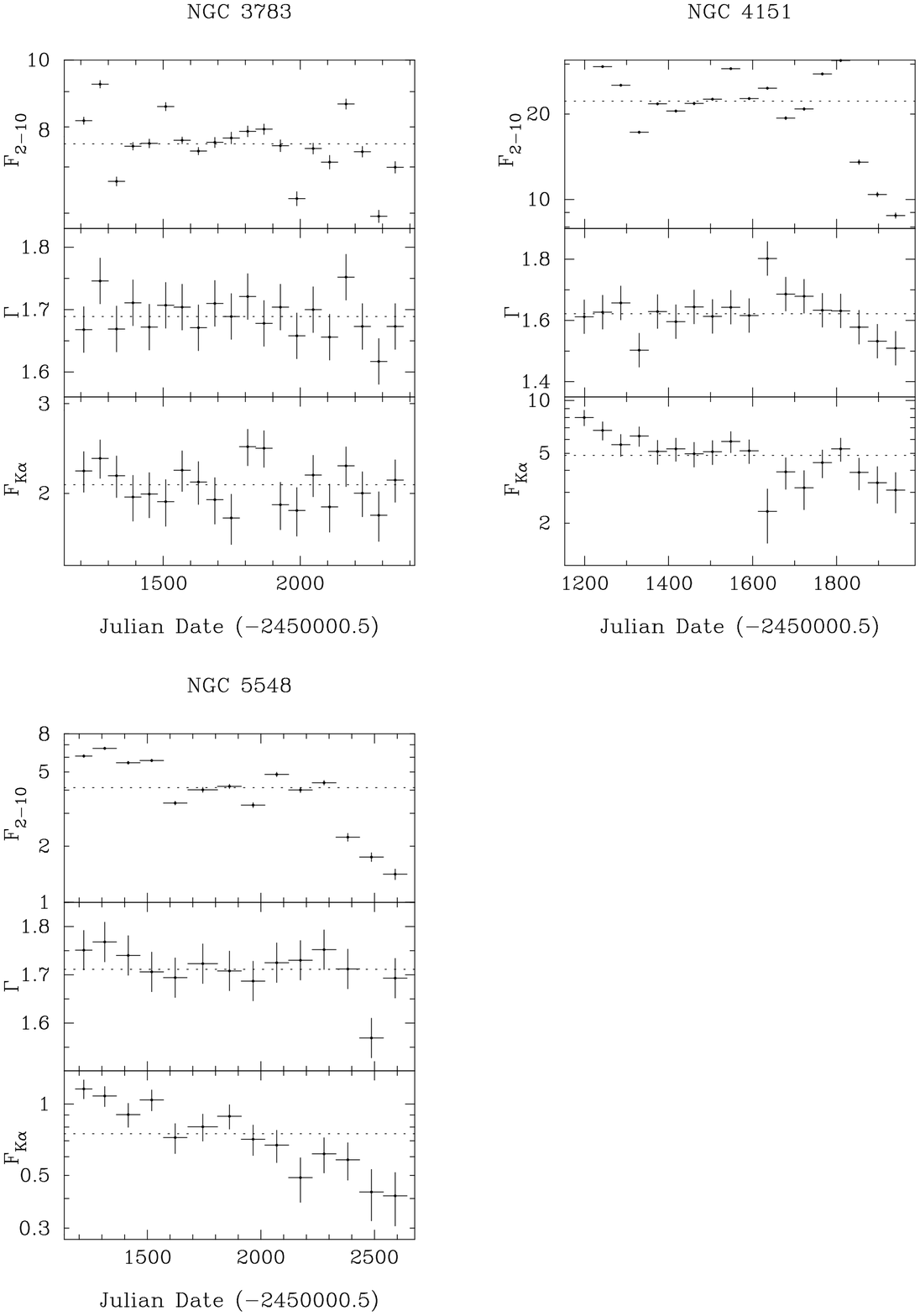}
\caption{Figure 2, cont'd.}
\end{figure}
\clearpage

\begin{figure}
\figurenum{3}
\epsscale{0.91}
\plotone{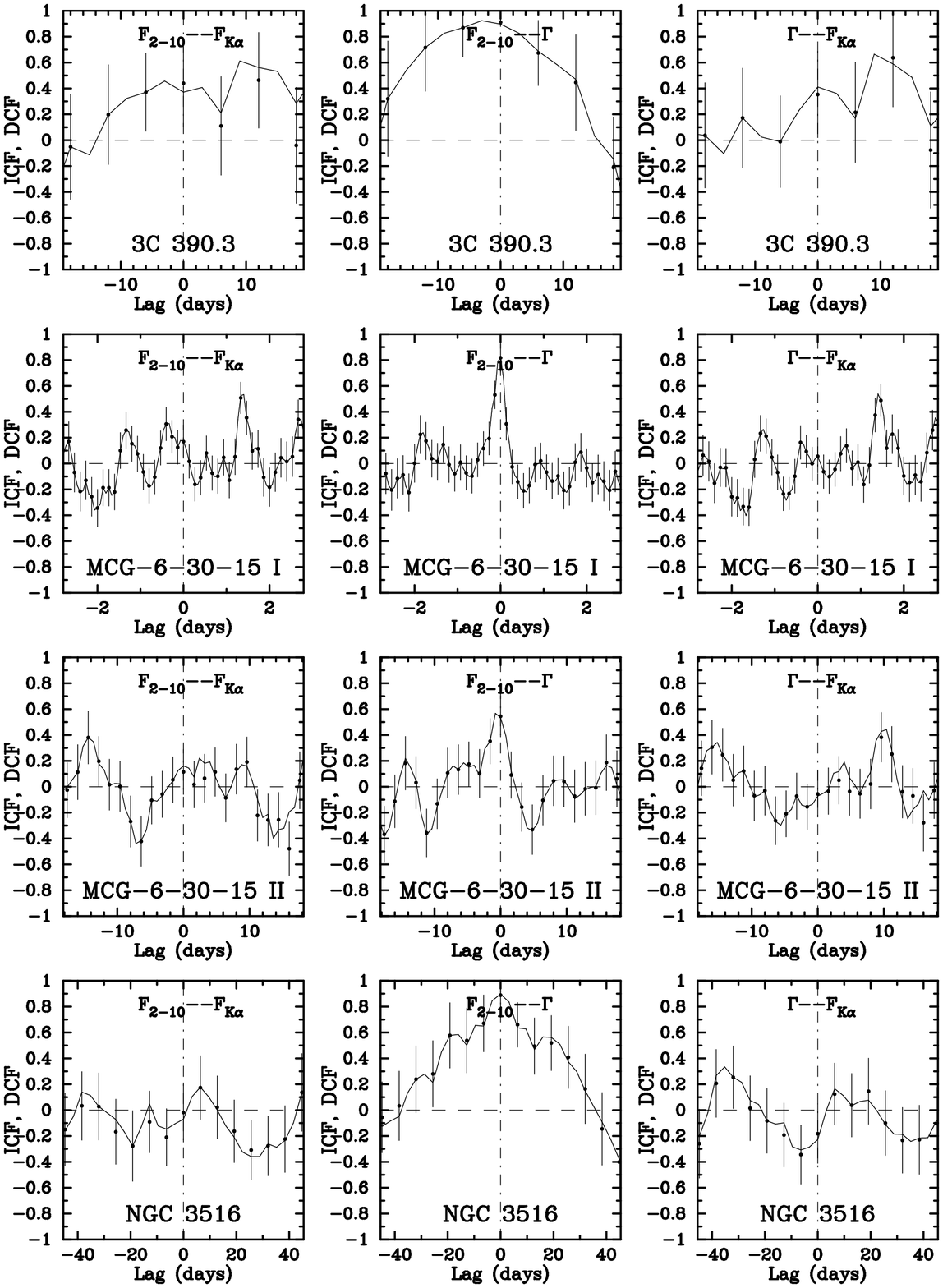}
\caption{Cross-correlation functions for the short time scale light curves. Lags are defined such that a positive lag means the first light curve leads the second.}
\end{figure}

\clearpage 

\setcounter{figure}{2}

\begin{figure}
\figurenum{3}
\epsscale{0.91}
\plotone{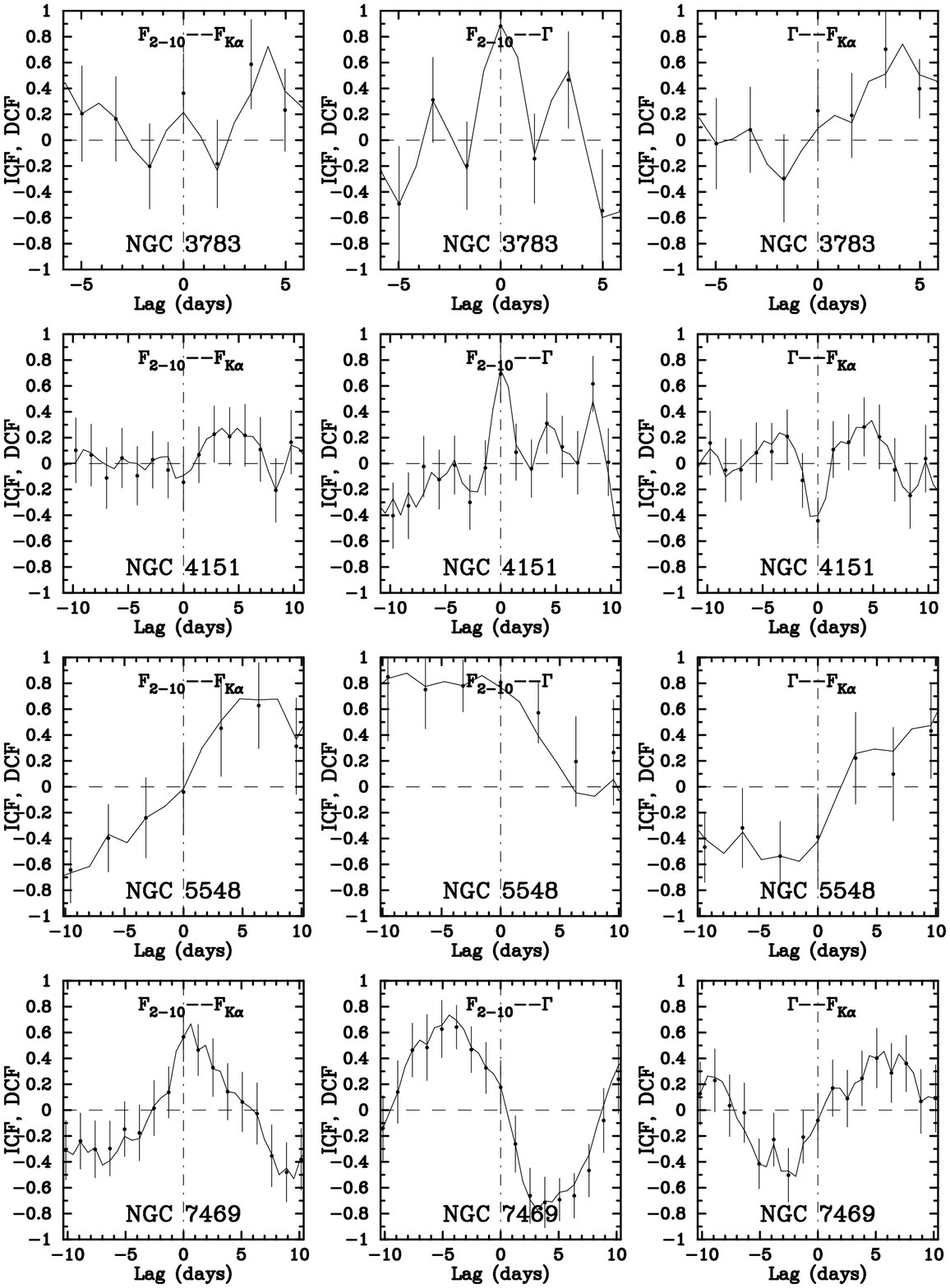}
\caption{Figure 3, cont'd}
\end{figure}

\clearpage

\begin{figure}
\figurenum{4}
\epsscale{0.91}
\plotone{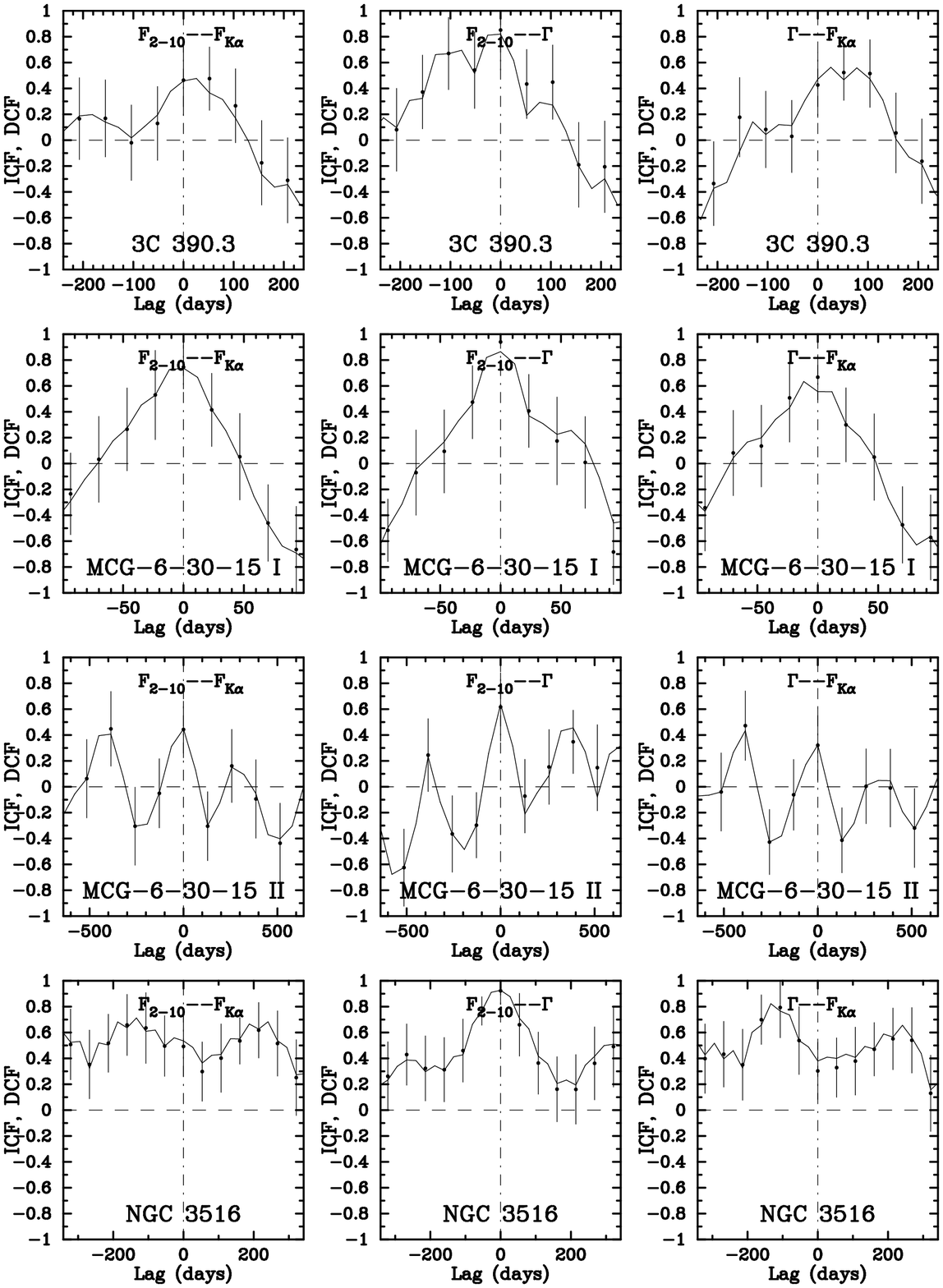}
\caption{Cross-correlation functions for the long time scale light curves. Lags are defined such that a positive lag means the first light curve leads the second.}
\end{figure}

\clearpage 

\setcounter{figure}{3}

\begin{figure}
\figurenum{4}
\epsscale{0.91}
\plotone{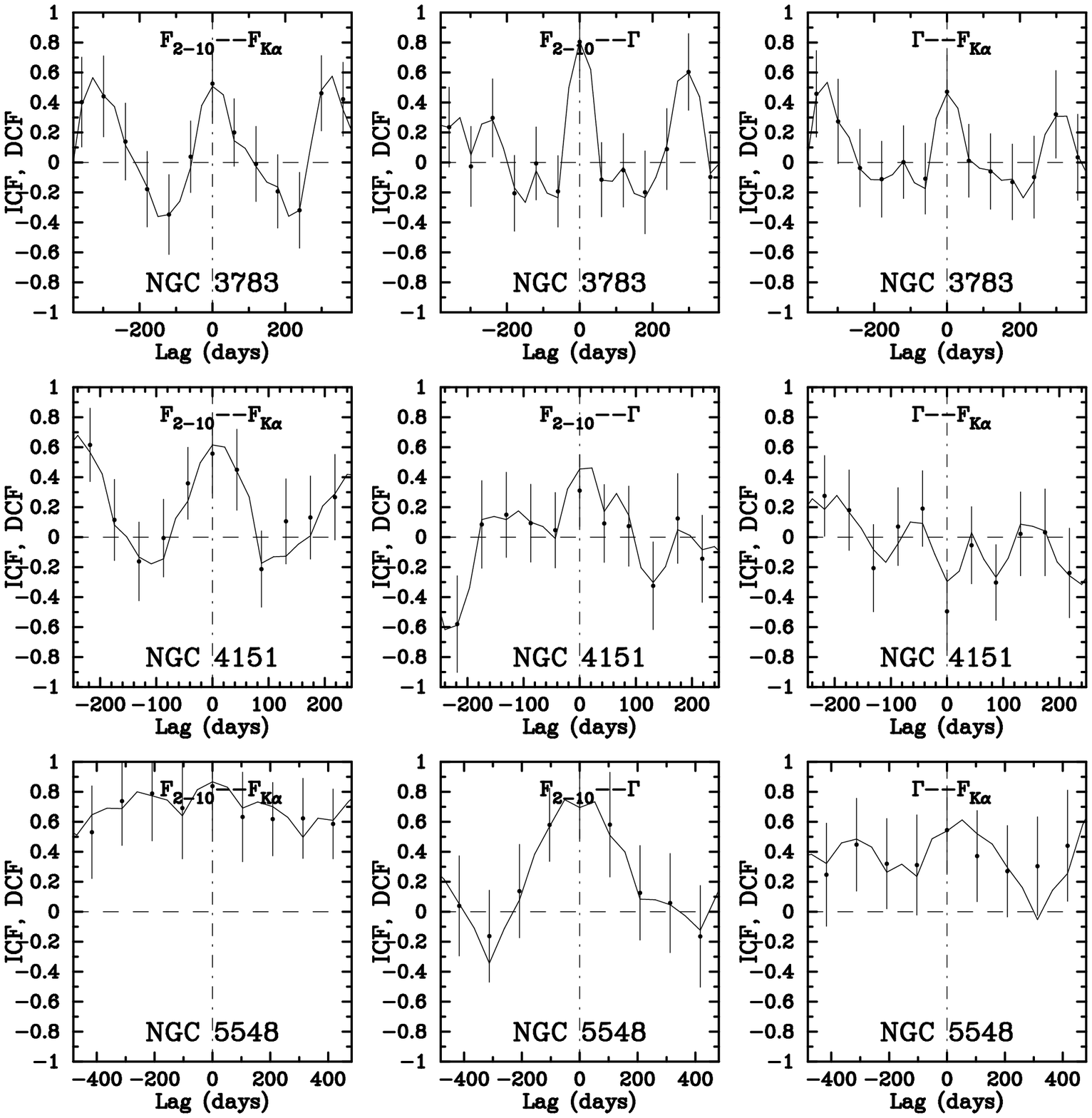}
\caption{Figure 4, cont'd}
\end{figure}
\clearpage

\begin{figure}
\figurenum{5}
\epsscale{0.94}
\plotone{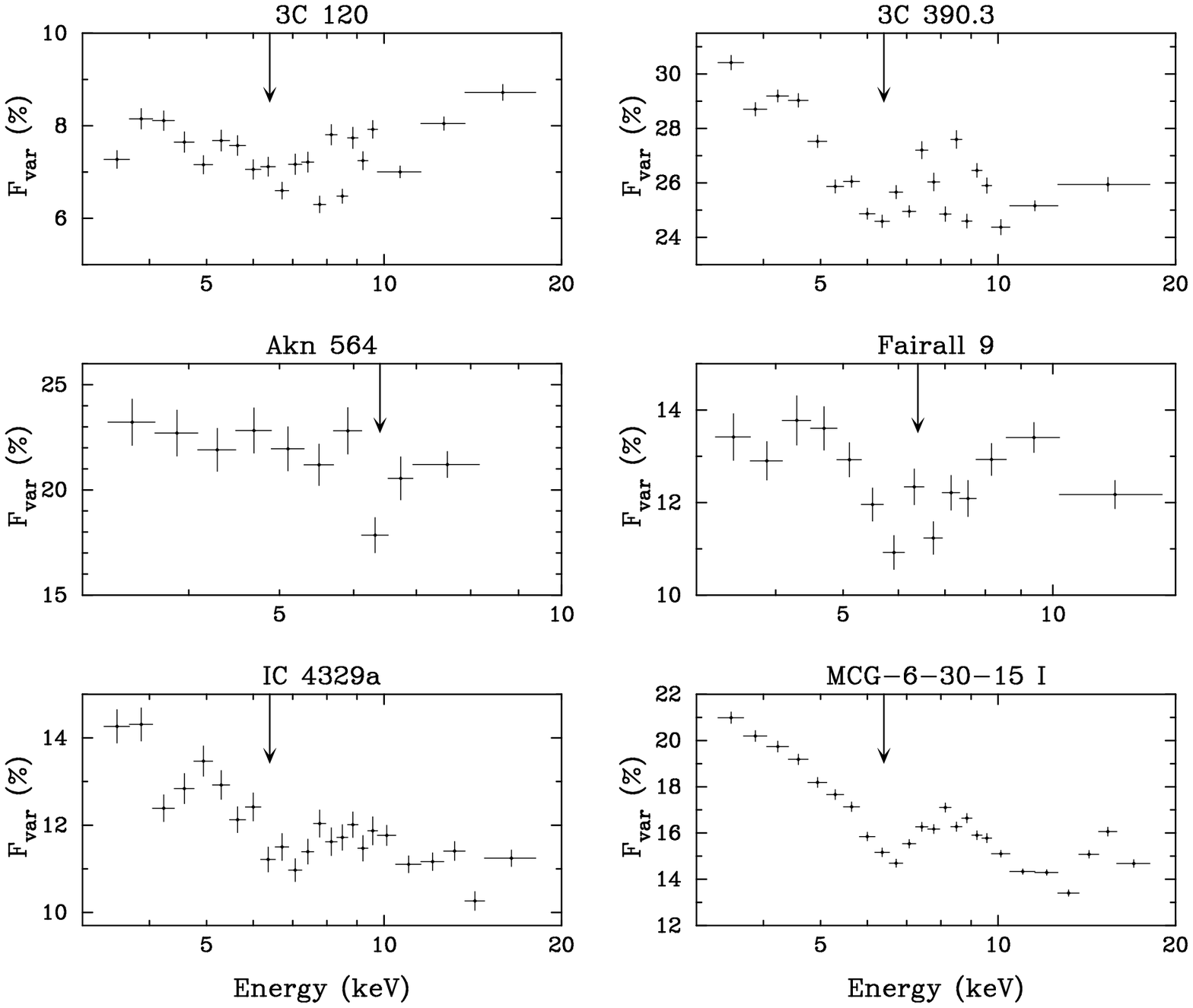}
\caption{Short time scale $\fv$ spectra. The downward arrows at 6.4~keV indicate the depression in broadband continuum variability
due to relatively reduced Fe \ka\ line variability.}
\end{figure}
\clearpage 
\setcounter{figure}{4}
\begin{figure}
\figurenum{5}
\epsscale{0.94}
\plotone{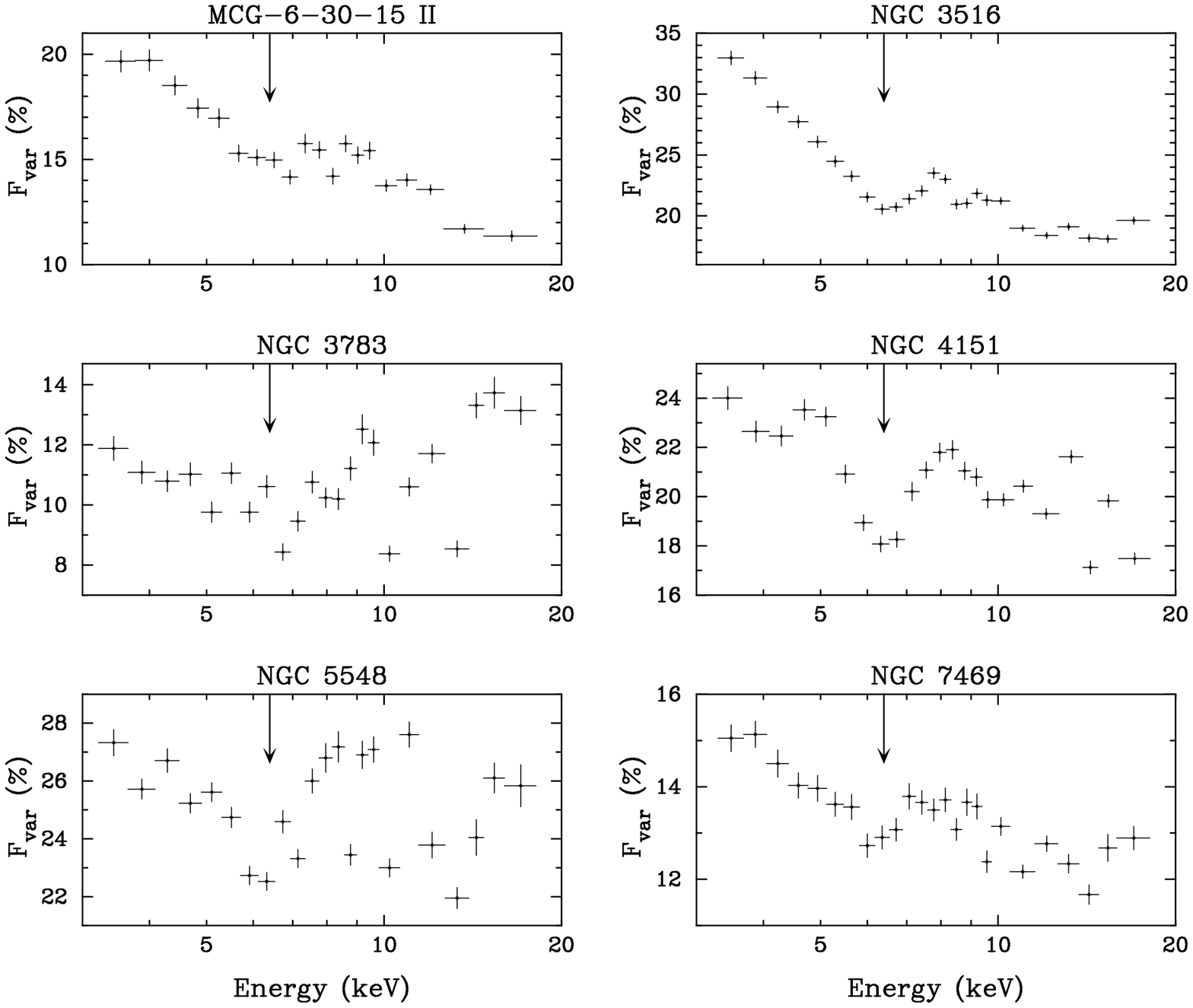}
\caption{Figure 5, cont'd}
\end{figure}
\clearpage

\begin{figure}
\figurenum{6}
\epsscale{0.94}
\plotone{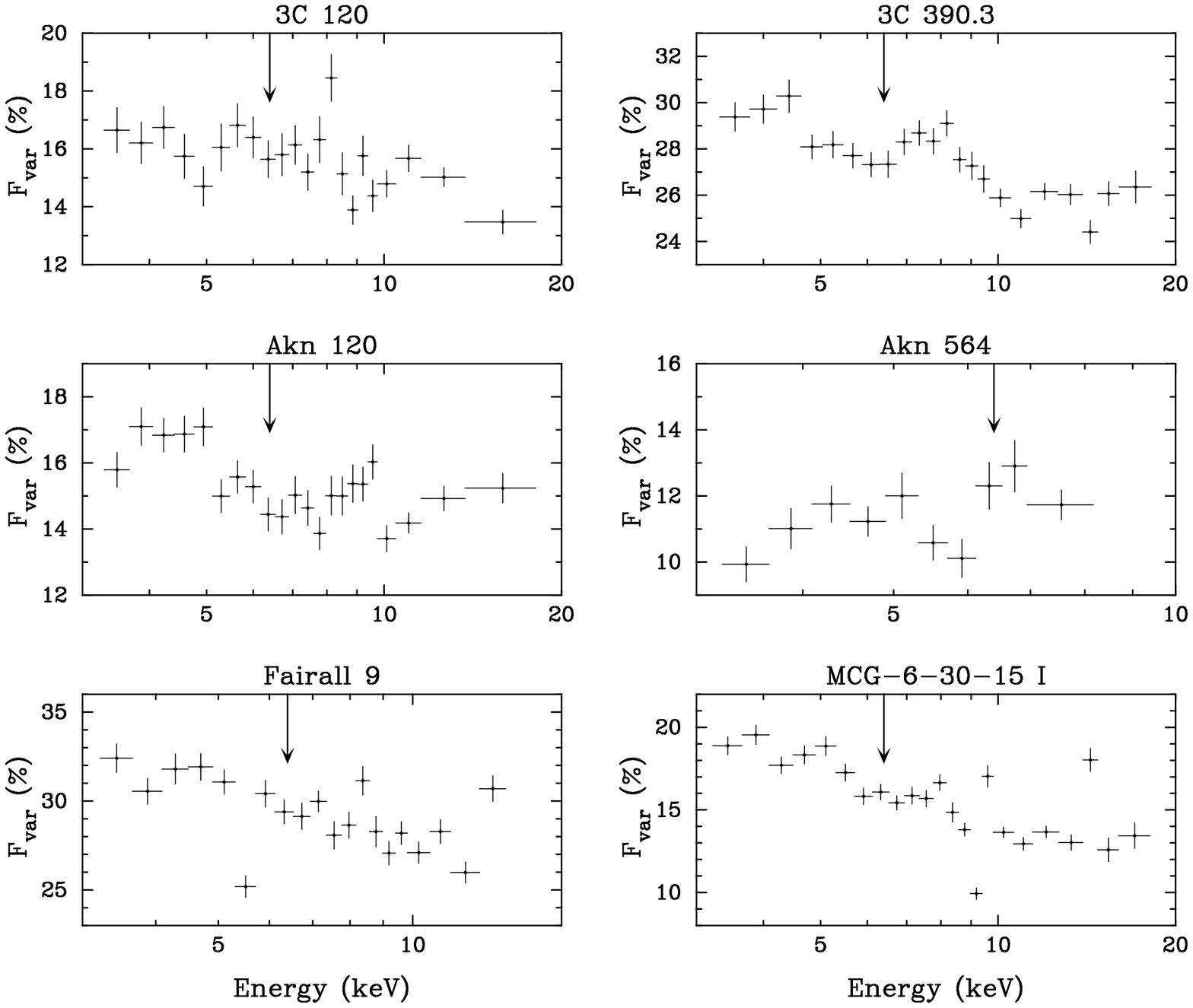}
\caption{Long time scale $\fv$ spectra. The downward arrows at 6.4~keV indicate the depression in broadband continuum variability
due to relatively reduced Fe \ka\ line variability.}
\end{figure}
\clearpage 
\setcounter{figure}{5}
\begin{figure}
\figurenum{6}
\epsscale{0.94}
\plotone{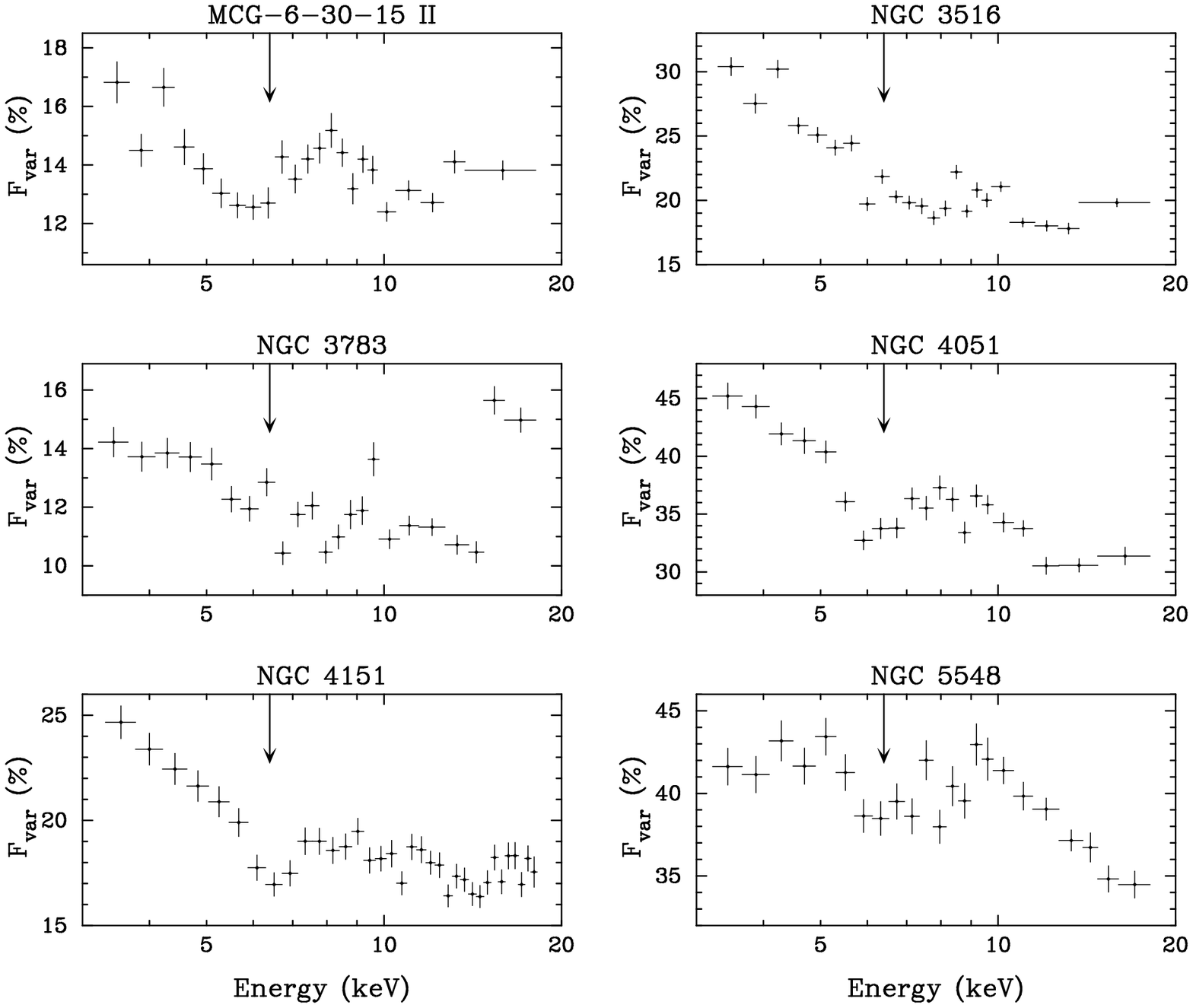}
\caption{Figure 6, cont'd}
\end{figure} 


\end{document}